\documentclass[showpacs,prd,reprint,preprintnumbers,amsmath,amssymb,nofootinbib]{revtex4-1}

\usepackage{graphicx,subfigure,multirow}
 \usepackage{longtable}
\usepackage{dcolumn}
\usepackage{bm}
\usepackage[all]{xy}
\usepackage{color}
\usepackage{slashed}
\usepackage[usenames,dvipsnames]{xcolor}
\usepackage[unicode=true,pdfusetitle,
 bookmarks=true,bookmarksnumbered=false,bookmarksopen=false,citecolor=Turquoise,
 breaklinks=false,pdfborder={0 0 1},backref=false,colorlinks=true]
 {hyperref}

 \usepackage{booktabs}
\usepackage{mathrsfs}
\usepackage{epsfig}
\usepackage{graphicx}
\usepackage{subfigure}
\usepackage{dcolumn}
\usepackage{bm}
\usepackage{amsmath}
\usepackage{slashed}
\usepackage{multirow}
\usepackage{color}
\usepackage{dcolumn}
\usepackage{bm}
\usepackage{color}
 \newcommand{\ud}{\mathrm{d}}

 \newcommand{\lsim}{{\;\raise0.3ex\hbox{$<$\kern-0.75em\raise-1.1ex\hbox{$\sim$}}\;}}
\newcommand{\gsim}{{\;\raise0.3ex\hbox{$>$\kern-0.75em\raise-1.1ex\hbox{$\sim$}}\;}}
\newcommand{\beq}{\begin{equation}}
\newcommand{\eeq}{\end{equation}}
\newcommand{\bea}{\begin{eqnarray}}
\newcommand{\eea}{\end{eqnarray}}

\newcommand{\pythia}{{\sc Pythia\,8\,}}

\def\baa{\begin{array}}
\def\eaa{\end{array}}

\mathchardef\minus="002D

\newcommand{\GeV}{\mathrm{\;GeV}}

\newcommand{\tabincell}[2]{\begin{tabular}{@{}#1@{}}#2\end{tabular}}

\begin{document}

\title{Tagging a jet from a dark sector with Jet-substructures at colliders}

\author{Myeonghun Park$^{2,3}$}
\email{parc.seoultech@seoultech.ac.kr}
\author{Mengchao Zhang$^{1,3}$}
\email{Corresponding author: mczhang@jnu.edu.cn}

\vspace{0.5cm}

\affiliation{$^1$Department of Physics, Jinan University, Guangzhou 510632, People’s Republic of China}
\affiliation{$^2$Institute of Convergence Fundamental Studies and School of Liberal Arts, Seoultech, Seoul 01811, Korea}
\affiliation{$^3$Center for Theoretical Physics of the Universe, Institute for Basic Science (IBS), Daejeon, 34051, Korea}

\vspace{0.5cm}

\date{\today}

\begin{abstract}
The phenomenology of dark sector is complicated if dark sector is charged under a confined hidden gauge group. 
In such kind of model, a dark parton produced at a high energy collider showers and hadronize to a cluster of dark mesons. 
Dark mesons then decay to visible particles and produce a jet-like signal, which is called ``dark jet" in this work.
Collider signal of dark jet depends on the property of dark mesons.
For example, a finite lifetime of dark meson would provide displaced vertex or displaced track, thus one can use these displaced objects to tag dark jet. 
However if the lifetime of dark meson is collider-negligible (too short to manifest a displaced vertex), it would be difficult to distinguish a dark jet from SM QCD jets.  
In this work we propose a new tagging strategy to identify dark jets from QCD backgrounds. 
This strategy is based on jet-substructure analysis. 
We study various jet-substructure variables and find out variables with good discrimination ability. 
Our result shows that by combining multiple jet-substructure variables, one could distinguish dark jets from QCD background, and thus enhance the sensitivity of dark sector search at collider. 

\end{abstract}



\maketitle

\section{Introduction}
\label{sec:introduction}

The existence of Dark Matter (DM) in our universe has been confirmed indirectly with its gravitational effects\,\cite{Ade:2015xua}. 
Still we have no idea about the nature of DM as we have not found DM ``directly" with various DM experiments. 
Especially, WIMP (Weakly Interacting Massive Particle) as the most popular DM paradigm has been a subject for various experiments including space indirect searches, nucleon scattering direct searches, and collider experiments.
However, we have excluded a wide range in the parameter space of WIMP from null results in above searches\,\cite{DirectSearch1,DirectSearch2,LHCSearch1,LHCSearch2,LHCSearch3,LHCSearch4,LHCSearch5}. 
In additional to the WIMP paradigm, another DM scenario called asymmetric DM\,\cite{Nussinov:1985xr,Kaplan:1991ah,Barr:1990ca,Barr:1991qn,Dodelson:1991iv,Fujii:2002aj,Kitano:2004sv,Farrar:2005zd,Kitano:2008tk,Gudnason:2006ug,Kaplan:2009ag,Shelton:2010ta,Davoudiasl:2010am,Buckley:2010ui,Cohen:2010kn,Frandsen:2011kt,Petraki:2013wwa,Zurek:2013wia,Ibe:2018juk,Ibe:2018tex}, which is motivated by the coincidence of the abundance of visible matter and DM as $\Omega_{DM} \simeq 5\Omega_B$, has attracted attentions. 
In asymmetric DM paradigm, DM and its antiparticle aDM(anti-Dark Matter) are produced not equally in the early universe. 
Subsequently, annihilation between DM and aDM eliminates aDM in the universe, and the remaining DM particles compose current relic density.

In order to stabilize DM and annihilate aDM efficiently, a hidden gauge group is generally introduced in asymmetric DM model. 
If some particle charged under this hidden gauge group can be produced at collider, then it is possible to study the dark sector through final state radiation of hidden gauge bosons. 
For example, if DM is charged under a $U(1)'$, then energetic DM produced at collider will radiate $U(1)'$ gauge boson(dark photon $\gamma'$). 
Dark photon decays back to the Standard Model (SM) particles through a kinetic mixing with SM photon, and leads to prompt/long-lived lepton jets or narrow jet signal at collider\,\cite{Bai:2015nfa,Buschmann:2015awa,Zhang:2016sll}.
If hidden gauge group in dark sector is $SU(N_d)$ which cause confinement at a certain scale $\Lambda_d$, 
then an energetic dark parton, which is the particle charged under $SU(N_d)$, will shower multiple dark partons then hadronize to a cluster of dark hadrons(most of them are the lightest dark meson). 
Dark hadrons then decay to particles in the SM, through some portals, and produce a jet-like signal. In this work we will call it ``dark jet". 

The property of dark jet depends on dark sector setting and the portal between dark sector and the SM. 
Previous phenomenology studies of different kinds of dark jets can be found in~\cite{Strassler:2006im,Strassler:2006ri,Strassler:2006qa,Han:2007ae,Verducci:2011zz,Chan:2011aa,Strassler:2008fv,Cohen:2015toa,Bai:2013xga,Schwaller:2015gea,Beauchesne:2017yhh,Pierce:2017taw,Beauchesne:2018myj,Renner:2018fhh}.
These studies generally rely on displaced objects from long-lived dark meson decay\footnote{In some scenario, lightest dark meson can decay to bottom quark pair, thus b-tagging can be utilized. But it requires dark mesons to be quite heavy. We don't consider this scenario in this work.}.
For example, in~\cite{Schwaller:2015gea} authors make dark meson long-lived by introducing a heavy mediator through which dark meson decays to SM quark pair. 
Thus there will be some displaced tracks inside dark jet, and one can utilize these displaced tracks to enhance collider search ability.
Recently, a new dark jet study based on a flavor structure in dark quark sector called semi-visible jet is proposed in\,\cite{Cohen:2015toa}. 
In their scenario, a certain amount of missing energy, which comes from some stable light dark hadrons, is collimated with dark jet, and a transverse mass of two leading jets in the final states becomes useful to discriminate dark jets pair signal from SM background.

\begin{figure}[t!]
\begin{center}
\begin{tabular}{cc}
\includegraphics[width=0.40\textwidth]{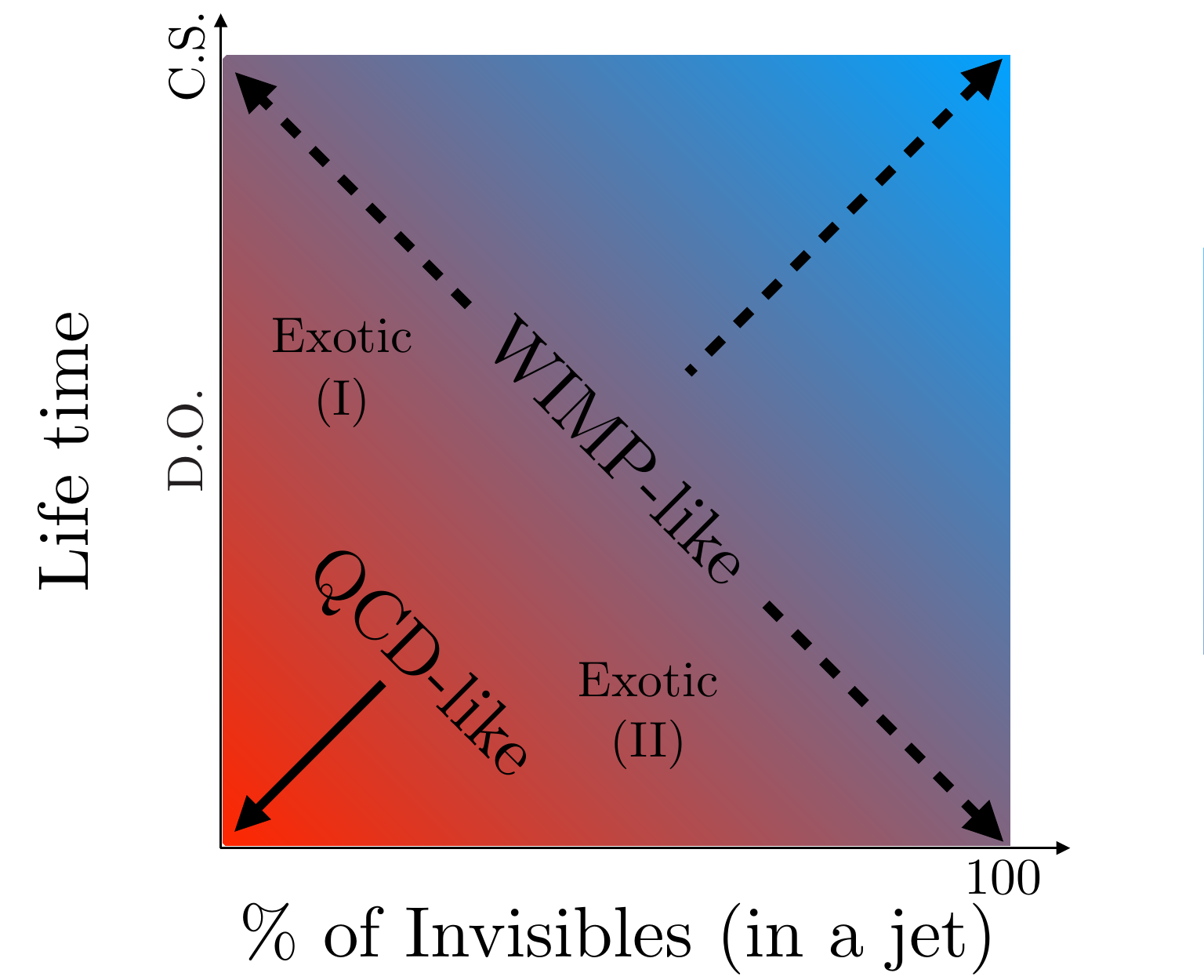}
\end{tabular}
\end{center}
\caption{\label{signal} We present a diagram to divide a jet-type from dark QCD in terms of (x-axis) percentage of stable (invisible) hadrons in a jet and (y-axis) life time of dark mesons. Here C.S. means a life time enough to be ``collider stable"  and D.O. stands for a sizable life time to be tagged with ``displaced objects". 
}
\end{figure}

All these search methods become noneffective if all or most of light dark mesons decay to SM particles promptly, because in this case these is no displaced objects or missing energy inside dark jet and dark jet looks like SM QCD jet. 
Inversely, if all or most of dark hadrons are collider stable, then there is only missing energy in final state. 
In this work we will not study this WIMP-like signal.
For illustration, in Fig.\,\ref{signal} we categorize dark jets signatures according to life-time of dark hadrons and the fraction of invisible particles inside a jet. Here we categorize dark jet searches into four; 
\begin{itemize}
\item Exotic (I): One can identify dark jet via displaced objects (D.O.) induced by lone-lived dark hadron decay. 
\item Exotic (II): Some stable dark hadrons (dark baryon and also some dark meson) occupy non-negligible portion of a  dark jet, which make various kinematic variables useful.
\item WIMP-like: Dark hadrons are stable or collider stable (C.S.). In this case the signal is like WIMP. 
\item SM QCD-like: There is no displaced objects or missing energy inside dark jet, and dark jet looks like a SM QCD jet.
\end{itemize}

In order to distinguish a ``SM QCD-like" dark jet from SM QCD background jet, we suggest to look inside a jet and utilize various jet-substructure techniques.
Due to recent improvements in quark-gluon jet discrimination with jet-substructure and corresponding applications in different New Physics searches\,\cite{Gallicchio:2010dq,Larkoski:2013eya,FerreiradeLima:2016gcz,Moult:2016cvt,Bhattacherjee:2016bpy}, we argue that we are at the stage of discriminating dark QCD jets from SM QCD jets. 
Jet-substructure reveals underlying structure of a QCD-like model, such as color factor or confinement scale. 
Thus it is possible for us to discriminate between a jet from dark QCD and a jet from SM QCD, provided dark QCD is not the same as SM QCD.

In next section we briefly introduce our model, and discuss which model setting can cause a ``SM QCD-like" dark jet.
Section 3 is dedicated to a comprehensive study of various jet-substructure variables and an exhibition of their discriminant performance.
In section 4 we use an example at LHC to show the effect of our dark jet tagging method. 
Then we summarize this work in section 5. 
A brief discussion of theoretical uncertainties from Monte Carlo simulation will be given in appendix A.
In appendix B we discuss the feasibility of our method.

\section{Benchmark scenarios for dark QCD models}

We introduce a new non-Abelian gauge group $SU(N_d)$ which describes dynamics in the dark sector.
Several light dark quarks as fundamental representations of $SU(N_d)$ are also required for constitution of dark hadron. 
Here, a light dark quark means a dark quark which is lighter than dark confinement scale $\Lambda_d$.
For $SU(N_d)$ to confine, the number of light dark quarks flavors $n_f$ should be smaller than $\frac{11}{2}N_d$.
At an energy scale much higher than $\Lambda_d$, the Lagrangian of dark sector can be written as:
\begin{equation}
\mathcal{L}_d = \bar{q'_i} (i\, \slashed D - m_{q'_i}) q'_i - \frac{1}{4}G'^{\mu\nu}G'_{\mu\nu},
\end{equation}
with $q'$ and $G'^{\mu\nu}$ denote dark quarks and dark gluon field strength respectively. $D_{\mu}$ corresponds to the covariant derivative of $SU(N_d)$, and $i$ is the flavor index of dark quarks. 
For minimality, we set the dark quarks to be SM singlet.
In order to produce dark partons at collider and decay dark hadrons, a mediator between dark sector and SM sector is required. 
It could be a bi-fundamental scalar $X$ which is charged under both $SU(N_d)$ and SM $SU(3)$\,\cite{Bai:2013xga}
\beq
\mathcal{L}_\text{med} = (D^{\mu} X)^{\dag}(D_{\mu} X) - M^2_{X}X^{\dag}X + \kappa_{ij} X\,\bar{q'_i} q_j +h.c. 
\eeq
or a heavy vector boson $Z'$ mediator connecting dark quark pair and SM quark pair\,\cite{Boveia:2016mrp}\,\footnote{Such a leptophobic $Z'$ will easily induce chiral anomaly\,\cite{Ellis:2017tkh,Ismail:2016tod}, but this topic is not so related to our present work, so we pay not attention to chiral anomaly tentatively.}
\beq
 \mathcal{L}_\text{med} = - \frac{1}{4}Z'^{\mu\nu}Z'_{\mu\nu} - \frac{1}{2}M^2_{Z'}Z'^{\mu}Z'_{\mu} + Z'_{\mu}(\bar{q'_i} \gamma^{\mu} q'_i + \bar{q_j} \gamma^{\mu} q_j )  \,.
\eeq
Here $q_i$ is SM quark, $i$ and $j$ are flavor index of dark quarks and SM quarks.
In order to make dark meson decay promptly, one can also extend hidden gauge group from $SU(N_d)$ to $SU(N_d)\times U(1)'$, with dark quark also charged under $U(1)'$~\cite{Knapen:2016hky}.
Thus light dark meson could decay to $\gamma'$ pair immediately, and $\gamma'$ decays to SM particles through kinetic mixing. 

The decay of dark hadrons depend on their spin, mass, and the mediator to the visible sector.
Here we give a comprehensive analysis to different kinds of dark hadrons and point out under which model setting the dark jet is "SM QCD-like".
\begin{itemize}
\item Dark pion $\pi_d$. 

Generally, spin-0 dark pions $\pi_d$ are much lighter then other dark hadrons, and they form a large part of particles in a dark jet until they decay. 
$\pi_d$ decays to quark pair through a high dimensional effective operator.
Due to a chiral flipping suppression, $\pi_d$ tends to decay to a heavy SM quark pair and its life time is closely related to the mass of the dark pion $m_{\pi_d}$. 
We use a formula from\,\cite{Schwaller:2015gea} to estimate the partial width of $\pi_d$ to a SM quark pair:
\begin{equation}
\Gamma(\pi_d \to q\bar{q}) = \frac{3 \kappa^4 f^2_{\pi_d} m^2_q}{32\pi M^4_{X}}m_{\pi_d}.
\end{equation}
Here $\kappa$ is the coupling among a mediator $X$, SM quark $q$ and a dark quark $q'$. $f_{\pi_d}$ is the decay constant of the dark pion, $m_q$ is the pole mass of the SM quarks and $M_{X}$ is the mass of the mediator $X$. 
$\kappa \sim 1$ is a natural choice. 
And approximately we have $m^2_{\pi_d} f^2_{\pi_d} \sim m_{q'} \Lambda^3_d$. 
Thus if $f_{\pi_d}$ and $m_{\pi_d}$ are several GeV, and $\pi_d$ decay channel to SM K-meson is open, then a $M_X$ around several hundred GeV can cause the proper decay length of $\pi_d$ to be shorter than 1mm, i.e. a promptly decaying dark pion.
This range for a mediator mass is still allowed by previous searches\,\cite{Schwaller:2015gea}.
If dark pion is even heavier and its decay channel to D-meson or even B-meson is open, the allowed parameter space for dark mesons decay promptly would be much larger. 

Another possibility is the case where there is an extra $U(1)'$ under which the dark quark is charged\,\cite{Knapen:2016hky}. In this case a dark pion will behave like SM pion and it decays to dark photon pair promptly.  A dark photon can decay into SM particles through a kinetic mixing with SM hyper charge $U(1)_\textrm{Y}$ where the kinetic mixing is parameterized by $\epsilon$. With current limits on dark photon\,\cite{Alexander:2016aln}, we find there are still huge surviving parameter space that can induce a prompt dark photon decay. For instance, a $0.4\GeV$ dark photon will decay promptly if $\epsilon \gtrsim 10^{-5}$. 
Thus it induces the prompt decay of a dark pion into SM particles.

In\,\cite{Cohen:2015toa} and a more recent paper\,\cite{Cohen:2017pzm}, authors consider a dark meson which is composed by different flavor dark quarks. 
In this case, a dark meson is stable and result in missing energy along with dark jet. 
While this dark flavor mixing meson can be un-stabilized by introducing some dark flavor violating portal. 
For example, interaction Lagrangian between two dark quark flavor and a mediator $X$ can be written as:
\begin{align}
\mathcal{L}_\text{int} = \kappa_{11} \bar{q'_1} q_1 X + \kappa_{21} \bar{q'_2} q_1 X + h.c.
\end{align}
Here $q_1$ is SM quark, $q'_1$ and $q'_2$ are two different flavor dark quarks. 
By integrating out the heavy mediator $X$, one can get an effective operator as\footnote{Fiertz transformation is used here. This kind of operator will not cause FCNCs(flavor-changing neutral currents) if we only consider one SM quark flavor.}:
\beq
\mathcal{L}_\text{eff} = \frac{\kappa_{11} \kappa^{\ast}_{21}}{M_X^2} (\bar{q'_1} \gamma_{\mu} q'_2) (\bar{q_1} \gamma^{\mu} q_1)\, + h.c.
\eeq
So depending on the parameters, flavor mixing dark meson $\pi_{d}$ ($q'_1\bar{q'_2}$ or $q'_2\bar{q'_1}$) could decay promptly into SM particles through this dark flavor violating operator. 

\item Dark rho meson $\rho_d$.

A dark rho meson is a spin-1 bound state made of dark quarks. 
Generally there is a mass splitting between a dark pion and a dark rho meson, which depends on the pole mass of dark quarks, because the mass of $m_{\pi_d}$ and $m_{\rho_d}$ can be approximated as:
\begin{eqnarray}
m^2_{\pi_d} \sim m_{q'} \Lambda_d \ , \   m^2_{\rho_d} \sim \Lambda^2_d
\end{eqnarray}
Thus if $m_{q'} \ll \Lambda_d$, a dark rho meson will decay promptly through decay channel $\rho_d \to \pi_d\, \pi_d$. 
If $m_{q'}$ is comparable with $\Lambda_d$, the mass splitting might not be enough to allow double dark pion decay. 
But due to the spin-1 property of $\rho_d$, its decay width will not be chiral-suppressed. 
Thus compared with $\pi_d$, it is much easier for $\rho_d$ to decay promptly.
Discussion of flavor mixing case is similar to dark pion, so we don't repeat it here. 

\item Dark baryon. 

The lightest dark baryon is stable and thus it can be a dark matter candidate. 
In $SU(3)_d$ case, the population ratio of baryons over mesons in a hadronization process is about $10\%$, which is negligible. 
If $N_d > 3$, the population ratio of baryon will be further suppressed. 
Only in $SU(2)_d$ case a considerable part of hadron in a dark jet consists of stable dark baryons. 
Thus in this work, we choose $N_d = 3$ and neglect dark baryons in dark hadronization process.

\item Dark glue-ball.

If all the dark quarks are much heavier than the confinement scale of $SU(N)_d$ ($m_{q'}\gg \Lambda_d$), the lightest dark hadron will be made of dark gluon. Thus one can call this dark hadron as a dark glue-ball. As a dark gluon and SM gluon belong to different gauge group, the decay of dark glue-ball is loop-induced by a heavy particles which have charges of both gauge groups. Thus the lifetime of dark glueball will be quite long. We will not discuss this scenario in this work.
\end{itemize}

Based on above discussion, we only consider dark meson $\pi_d$ and $\rho_d$ in our simulation, and we let all the dark mesons decay promptly in event generator. Detailed benchmark setting are listed in Tab.~\ref{Models}. 
In order to cover the diversity of dark QCD models, we consider different confinement scale, spectrum, and decay channels. 
Due to the non-perturbative nature of a QCD-like theory, some of those parameters need to be given by hands.  
In Tab.~\ref{Models}, constituent quark mass $\tilde{m}_{q'}$ is used to estimate dark hadron spectrum. 
And for simplicity, we assume $\pi_d$ and $\rho_d$ composed by different flavor dark quarks have identical mass and decay channel. 
In next section we will show how one can utilize jet sub-structure variables to distinguish a dark jet from SM QCD jets. 

\begin{table*}[t!]
\begin{center}\begin{tabular}{|c|c|c|c|c|c|c|p{2.7cm}|p{3cm}|c|}
\hline       &
 \multirow{ 2}{*}{$N_d$} & \multirow{ 2}{*}{$n_f$} & $\Lambda_d$ &  $\tilde{m}_{q'}$ & $m_{\pi_d}$ & $m_{\rho_d}$ & 
 \multirow{ 2}{*}{$\pi_d$ Decay Mode} & \multirow{ 2}{*}{$\rho_d$ Decay Mode} \\
  &  & & (GeV) &  (GeV) & (GeV) & (GeV) &  & \\
\hline$A$ &3&2&15&  20  &10&50&$\pi_d \to c \bar{c}$&$\rho_d \to \pi_d \pi_d$ \\
\hline$B$ &3&6&2&  2 &2&4.67&$\pi_d \to s \bar{s}$&$\rho_d \to \pi_d \pi_d$  \\
\hline \multirow{ 2}{*}{$C$} &\multirow{ 2}{*}{3}&\multirow{ 2}{*}{2}&\multirow{ 2}{*}{15}& \multirow{ 2}{*}{20} &\multirow{ 2}{*}{10}&\multirow{ 2}{*}{50}&$\pi_d \to \gamma' \gamma' $ {with $m_{\gamma'} = 4.0\GeV$} &\multirow{ 2}{*}{$\rho_d \to \pi_d \pi_d$}  \\
\hline \multirow{ 2}{*}{$D$} &\multirow{ 2}{*}{3}&\multirow{ 2}{*}{6}&\multirow{ 2}{*}{2}&\multirow{ 2}{*}{2} &\multirow{ 2}{*}{2}&\multirow{ 2}{*}{4.67}&$\pi_d \to \gamma' \gamma' $ {with $m_{\gamma'} = 0.7\GeV$} &\multirow{ 2}{*}{$\rho_d \to \pi_d \pi_d$} \\
\hline\end{tabular}
\caption{Benchmark models we considered in this work. All dark mesons are assumed to decay promptly. 
We mainly consider 2 cases: high $\Lambda_d$ case like A and C, and low $\Lambda_d$ case like B and D.
Parameters in a dark sector for A and C,  B and D are the same except the decay channel of a dark pion $\pi_d$.
$\pi_d$ and $\rho_d$ mass obey following two equations: $m_{\pi_d} = 2\tilde{m}_{q'} -\frac{3}{4}\frac{\Omega}{\tilde{m}^2_{q'}} $ and $m_{\rho_d} = 2\tilde{m}_{q'} +\frac{1}{4}\frac{\Omega}{\tilde{m}^2_{q'}} $ ~\cite{Griffiths:2008zz}, where $\Omega$ is proportional to binding energy.
Thus $m_{\rho_d}$ can be determined by constituent dark quark mass $\tilde{m}_{q'}$ and $m_{\pi_d}$.
The ratio of $\pi_d$ to $\rho_d$ after fragmentation is decided by $\frac{ \# \pi_d}{ \# \rho_d} = \frac{1}{3}e^{-\frac{m_{\pi_d}-m_{\rho_d}}{\Lambda_d}}$.
The branching ratios of their decay modes shown here are all $100\%$, if we don't give a specific value.
Decay modes of a dark photon $\gamma'$ with different mass can be found in \cite{Buschmann:2015awa}.}
\label{Models}
\end{center}
\end{table*}

\section{Jet-substructure Variables Analysis}

\begin{figure}[t!]
\begin{center}
\includegraphics[width=0.40\textwidth]{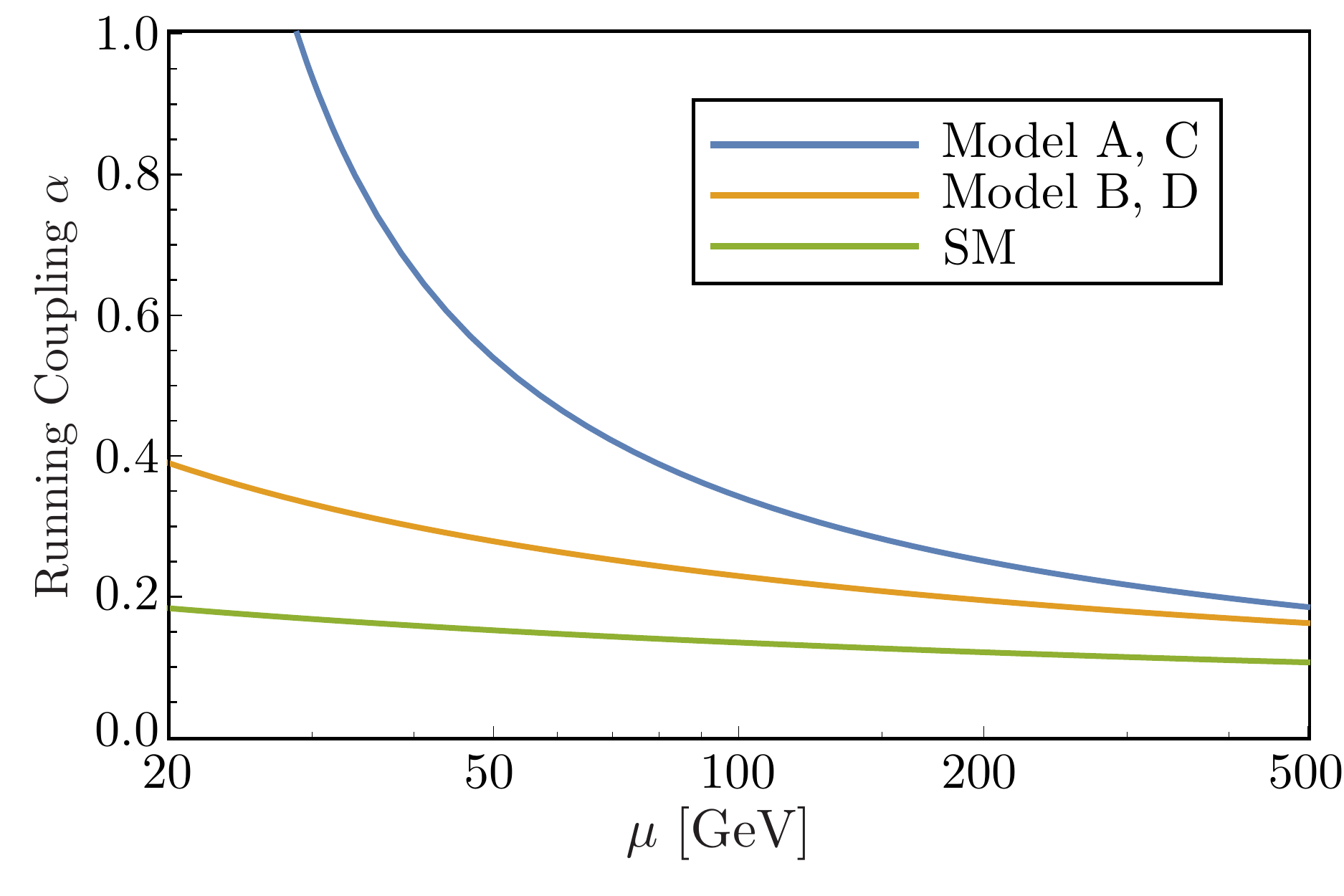}
\caption{QCD coupling running in dark sector and SM QCD. Definition of A(C) and B(D) can be found in Tab.\,I. For SM QCD we choose $n_f = 5$ as we don't include the effects from a top-quark.}
\label{coupling}
\end{center}
\end{figure}

Underlying parameters in a dark sector will affect the collider phenomenology of a dark jet. 
The RGE running of a dark sector gauge coupling $\alpha_d(\mu)$ is controlled by these parameters:
\begin{eqnarray}
\frac{\ud}{\ud \ln \mu^2} \alpha^{-1}_d(\mu) = \frac{1}{12 \pi} (11N_d - 2n_f),
\end{eqnarray}
with boundary condition $\alpha^{-1}_d(\Lambda_d) = 0$.
A comparison in a running coupling between SM QCD and various dark QCD models is shown in Fig.\,\ref{coupling} (Corresponding dark sector setting can be found in Tab.~\ref{Models}). 
Running coupling determines parton shower, which happens at a short distance smaller than $1/\Lambda_d$.
Then those showered partons fragment to dark hadrons. Finally dark hadrons decay back to SM particles which are measured by a detector.
Combining these three processes, the detector level measurements of jet-substructure variables, like jet mass or track multiplicity for a dark jet could be quite different from the expectations for SM QCD jets.

Dark jet originated from a single dark parton can be considered as a 1-prong jet.
Thus jet grooming\,\cite{Ellis:2009me,Ellis:2009su,Krohn:2009th} methods including mass dropping algorithm or pruning, which are suitable for reconstructing a boosted heavy object like a gauge boson ($W/Z/H$) or top-quark, are not expected to be effective in tagging a dark jet. 
Compared to 2 or 3-prong jet tagging, 1-prong jet tagging is easier due to a simpler jet structure. 
Jet-substructure variables used to tag a 1-prong jet roughly fall into two categories, infrared collinear\,(IRC) safe ones and IRC unsafe ones.

An IRC safe variable is not sensitive to soft or collinear radiations inside jet, i.e., contributions from extra radiation to an IRC safe variable are negligible if radiations are soft or collinear. 
Thus an analytical description of IRC safe variables is possible. 
We choose jet mass, two-points energy correlation function $C_1^{(\beta)}$\,\cite{Larkoski:2013eya} , and linear radial geometric moment (Girth)\,\cite{Gallicchio:2011xq} as our IRC safe variables.
Analytical descriptions of these IRC safe variables have been given in the literature, it helps us to understand our results which are mainly based on Monte Carlo simulation.

An IRC unsafe variable, for example the charged track multiplicity, is sensitive to soft and collinear radiations. 
Besides that, some IRC unsafe variables are also related to dark meson decay channel.
For those variables we will provide Monte Carlo based results and give some qualitative arguments.

We choose \pythia\cite{Sjostrand:2014zea} for shower and hadronization simulation. 
It has been shown that jet substructures obtained from \pythia fits well with real data \,\cite{ATLAS:2012am,Aad:2013gja}. 
Hidden Valley model\,\cite{Strassler:2006im} implemented in \pythia can be used to simulate dark QCD process, and recently the running of dark gauge coupling have been added to \pythia which greatly enhance the reliability of dark QCD simulation.
We generate three processes at the LHC; $f\bar{f} \to Z' \to q'\bar{q'}$, $qg \to Z q$, and $q\bar{q} \to Z g$ to study dark jet, quark jet, and gluon jet respectively.
For realistic analyses, we perform analyses at the detector level with {\sc DELPHES\,3}\,\cite{deFavereau:2013fsa}.
We use {\sc Fastjet\,}\cite{Cacciari:2011ma} to cluster final state particles with an anti-kt algorithm\,\cite{Cacciari:2008gp}.
The objects for a jet clustering are energy deposits in an electric calorimeter, a hadronic calorimeter and muons without isolation criterion. Because there can be a fraction of dark jet energy carried by muon, depending on the decay channel of dark pion\footnote{In this case, some dedicated method can be designed to tag a muon rich jet. But as we study the behavior of general jet substructure variables to cover various types of a dark jet, we will not pay special attention to muons in this work.}. 
Examining the discrimination performance of jet substructure variables with different choices of jet radius ($R$), jet transverse momentum ($p_T$), and jet algorithms can be interesting. In our study, we choose $R=0.4$ as it is a typical jet radius in the LHC experiment analyses for QCD jet and this choice was studied in the ATLAS light-quark and gluon jet discrimination\,\cite{Aad:2014gea}.
For the choice of jet transverse momentum $p_T$, we start with the range of $p_T\in (180\GeV,220\GeV)$ as this $p_T$ range has the minimum systematic uncertainties\,\cite{Aaboud:2017jcu} and it overlaps with the $p_T$ range in the ATLAS jet discrimination study\,\cite{Aad:2014gea}. We consider a detector geometry of pseudo rapidity $\eta\,\in(-2.5,2.5)$.

\begin{figure*}[t!]
  \centering
  \includegraphics[width=0.80\textwidth]{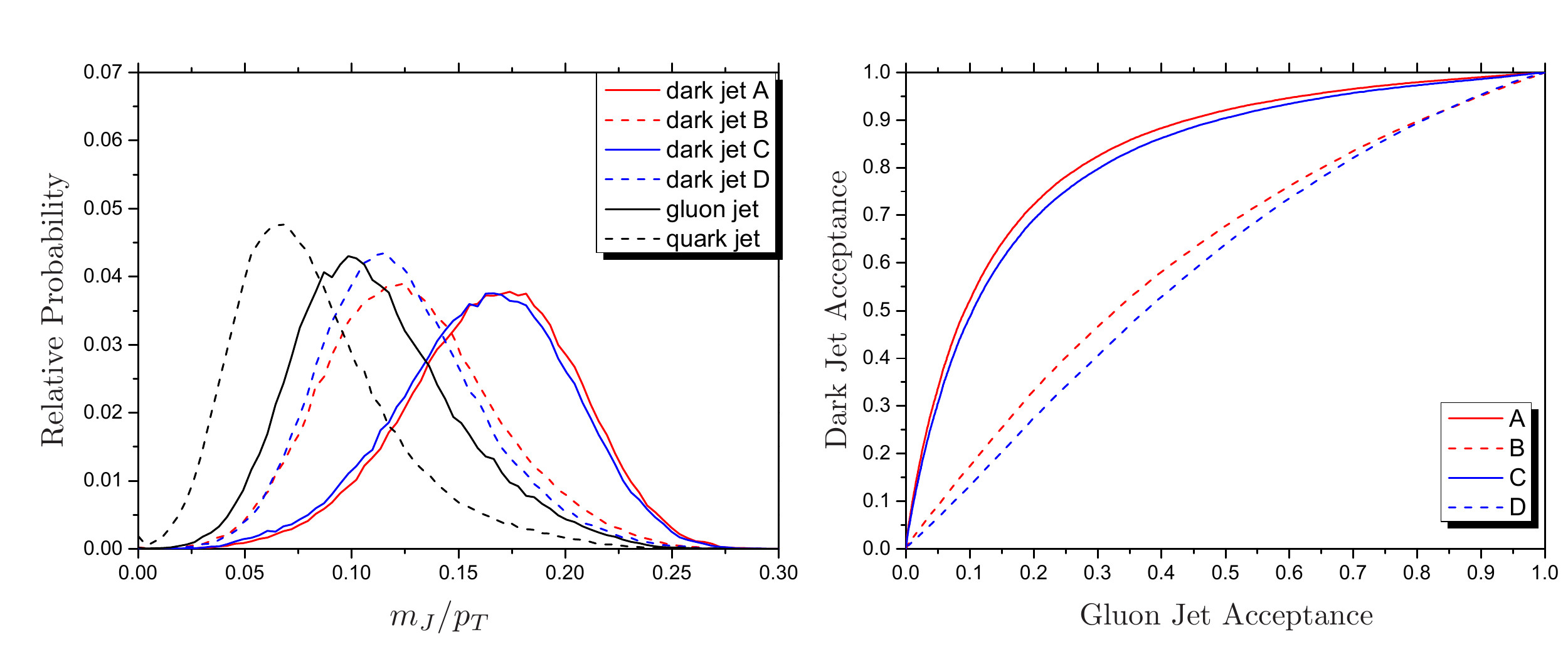}
\caption{Left: mass over $p_T$ distribution of dark jet, quark jet, and gluon jet with $ p_T \in (180\GeV,\,220\GeV)$. Right: ROC curve with $m_J/p_T$ used for the separation between a jet from various dark QCD models and SM gluon-initiated jet.}
\label{mass}
\end{figure*}

\subsection{Jet mass}
Jet mass, as a simple and intuitive variable which reflects the underlying structure of a jet, has been studied by decades\,\cite{Catani:1992ua,Ellwanger:1980uf,Kelley:2011tj,Li:2011hy,Li:2012bw,Chien:2012ur}. 
Jet mass originates from the virtuality of the primordial parton of a jet. 
As we consider the first order splitting process, a normalized differential cross section of virtuality is:
\begin{equation}
\frac{1}{\sigma} \frac{\ud \sigma}{\ud p^2} = \frac{C}{2\pi p^2} \int^{1-\epsilon}_{\epsilon} \ud z\,  \alpha(zp^2) P(z,p^2)\, ,
\end{equation}
where $\sigma = \int (\ud \sigma/\ud p^2)\ud p^2$ is the integrated jet cross section, $C$ is color factor, $p$ is the 4-momentum of a primordial parton and $p^2$ is its virtuality. $\epsilon$ is an infrared cut, $z$ is the energy fraction carried by a radiated parton, $\alpha(\mu)$ and $P(z,p^2)$ are QCD running coupling and splitting kernel respectively. 
Above fixed order result is divergent when a jet mass becomes zero, which is in conflict with experiment data.
In order to get a reasonable distribution, one needs to resum higher order corrections. 
In Leading Log order, differential cross section becomes: 
\begin{equation}
\frac{1}{\sigma} \frac{\ud \sigma}{\ud p^2} = \frac{\ud}{\ud p^2} S(p^2,Q^2),
\label{equa_mass1}
\end{equation}
which is a differential to the Sudakov factor $S(p^2,Q^2)$:
\begin{equation}
S(p^2,Q^2) = \exp\bigg\{-\int^{Q^2}_{p^2}\ud k^2 \frac{C}{2\pi k^2} \int^{1-\epsilon}_{\epsilon} \ud z\,  \alpha(zk^2) P(z,k^2)  \bigg\}\, .
\label{equa_mass2}
\end{equation}
Here $Q$ is the energy scale of corresponding hard process. 
This leading order result can roughly reproduce shape of the real data distribution from the LHC experiments.
Obviously, this distribution is determined by running coupling $\alpha(\mu)$ and color factor $C$. 
In order to get an intuition for jet mass distributions, we approximate Eq.\,(\ref{equa_mass1},\,\ref{equa_mass2}) below.
With fixing running $\alpha(\mu)$ as $\alpha$, $P(z,k^2)=1/z$, and choosing $\epsilon=p^2/Q^2$, we obtain the following approximation:
\beq
\frac{1}{\sigma_0} \frac{\ud \sigma}{\ud (p^2/Q^2)}  \approx  
\frac{C \alpha}{\pi} \frac{Q^2}{p^2} \ln\frac{Q^2}{p^2} \exp\bigg\{-\frac{C \alpha}{2\pi} \left(\ln\frac{Q^2}{p^2}\right)^2 \bigg\} .
\label{eq:massCA}
\eeq
As we see in the above eq.\,(\ref{eq:massCA}), the peak of a jet mass distribution moves to a right side as $C\alpha$ becomes lager.
Thus the peak of a jet mass distribution for gluon-initiated jet is on the right side compared to the peak of a distribution from a quark-initiate jet, as color factor for a gluon $C_A = 3$ is larger than the color factor $C_F = 4/3$ of a quark as in Fig.\,\ref{mass}. 

In SM QCD, the only difference between quark jet and gluon jet is color factor $C_F$ (for a quark) and $C_A$ (for a gluon). 
Even so, a dimensionless parameter $m_J/p_T$, jet mass divided by its $p_T$, is a good variable used in quark/gluon jet discrimination.
For a dark jet, because of a quite different running coupling and a possible different color factor, one could certainly expect a very different distribution of a jet mass compared to the case of SM QCD.

With considering subleading contributions, one can include the effect of a jet size or a hadronization\,\cite{Kelley:2011tj,Chien:2012ur}. 
In our study, we will not go further analytically, but utilize Monte Carlo simulation (\pythia) to get numerical results.
Jet mass distributions from different models in Tab.\ref{Models} and SM QCD are shown in Fig.\,\ref{mass}.

As the gauge coupling strength of a dark QCD model A (C) is larger than the gauge coupling strength of B (D), a jet from model A or C has larger mass than a jet from model B or D. Equivalently, a dark QCD model with a higher confinement scale $\Lambda_d$ is easier to be distinguished from SM QCD.
We can check discrimination performance with ROC (receiver operating characteristic) curves in the right column of Fig.\,\ref{mass}.
We also argue that a jet mass is not sensitive to final states (SM particles from the decay of dark mesons) as jet mass distributions of A(B) almost overlaps the distribution of C(D) in Fig.\,\ref{mass}. 

\subsection{Two points energy correlation function $C_1^{(\beta)}$}
Another variable which is useful to probe properties of an one-prong jet is two-points energy correlation function\,\cite{Larkoski:2013eya}:
\begin{equation}
C_1^{(\beta)} =   \displaystyle{\sum_{i < j \in J}} z_i z_j (R_{ij})^{\beta} ,
\end{equation}
with $z_i = {{p_T}_i}/{ \displaystyle{\sum_{i \in J} {p_T}_i}} $ is the $p_T$ fraction carried by component $i$ within a jet $J$, and $R_{ij}$ is the distance between component $i$ and  $j$. 
As studied in \cite{Larkoski:2013eya}, the advantage of infrared collinear safe variable like $C_1^{(\beta)}$ is that analytical calculation of them is possible. Here we adopt analytical results from \cite{Larkoski:2013eya} to see the dependence of $C_1^{(\beta)}$ on the parameters of dark QCD.
Firstly one can consider the simplest case, which is the fixed leading order distribution with treating $\alpha$ as a constant for simplicity;
\beq
\frac{1}{\sigma} \frac{\ud \sigma}{\ud C_1^{(\beta)}} =  \frac{2\alpha C}{\pi}  \int^{R_0}_0 \frac{\ud \theta}{\theta}  
\int^1_0 \frac{\ud z}{z} \delta\left(z(1-z) \theta^{\beta} - C_1^{(\beta)} \right).
\eeq
Here $R_0$ is the size of a jet, which is the upper limit of a splitting angle in shower process. 
After integrations, we get
\bea
\frac{1}{\sigma} \frac{d \sigma}{d C_1^{(\beta)}} &=&  \frac{2\alpha C}{\pi \beta C_1^{(\beta)} } \ln \left( \frac{1 + \sqrt{1-4C_1^{(\beta)}/R_0^{\beta}}}{1 - \sqrt{1-4C_1^{(\beta)}/R_0^{\beta}} }\right)  \nonumber \\
&\equiv& \frac{2\alpha C}{\pi \beta C_1^{(\beta)} }\, L(C_1^{(\beta)}/R_0^{\beta})\, .
\eea
Similar to our previous fixed order calculation for the distribution of a jet mass, $C_1^{(\beta)}$ distribution is also divergent in the soft and collinear region.
With a leading order resummation, one obtains:
\begin{eqnarray}
&&\frac{1}{\sigma} \frac{\ud \sigma}{\ud C_1^{(\beta)}} = \frac{\ud}{\ud C_1^{(\beta)}} \exp\left( -\int^{R_0^{\beta}}_{C_1^{(\beta)}} \ud \tilde{C}\, \frac{2\alpha C}{\pi \beta \tilde{C} } L(\tilde{C}/R_0^{\beta})  \right)  \nonumber \\
&&= \frac{2\alpha C}{\pi \beta C_1^{(\beta)} } L(C_1^{(\beta)}/R_0^{\beta}) \exp\left( -\int^{R_0^{\beta}}_{C_1^{(\beta)}}\ud \tilde{C}\,  \frac{2\alpha C}{\pi \beta \tilde{C} } L(\tilde{C}/R_0^{\beta})  \right)\, .~~~~
\label{eq_C1}
\end{eqnarray}
One can notice that the probability in soft and collinear region will be suppressed by an exponent. 
As we have seen in jet mass distribution, the peak value of dark jet $C_1^{(\beta)}$ distribution is larger than the peak value of SM QCD jet $C_1^{(\beta)}$ distribution, as dark QCD has a larger coupling compared to SM QCD.

\begin{figure*}[t!]
\centering
\includegraphics[width=15cm]{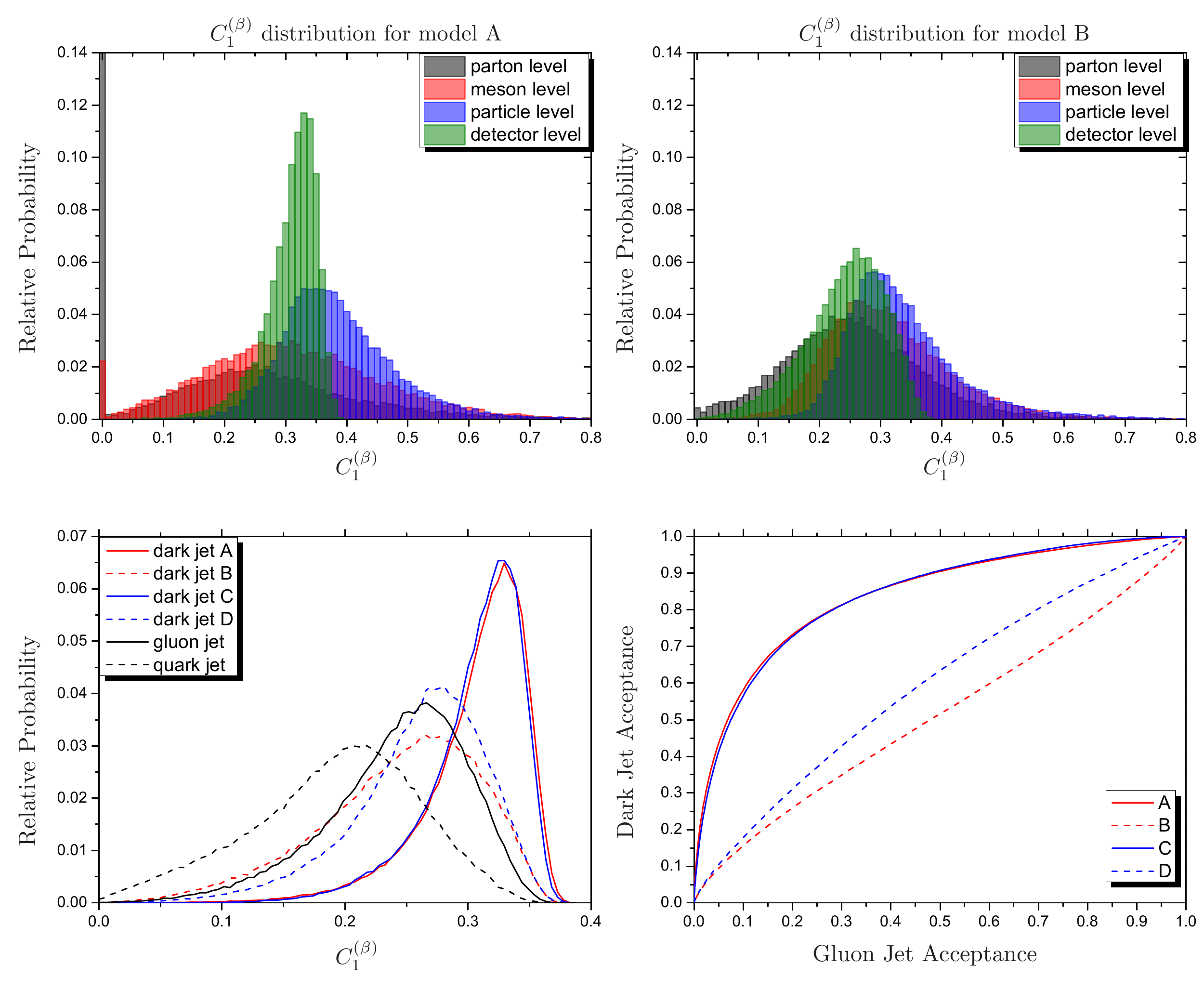}
\caption{Top left: $C_1^{(\beta)}$ distribution of dark jet with $p_T \in (180\GeV,\,220\GeV)$ at parton level,  meson level, final state particle level, and detector level for dark QCD model A (corresponding to a high dark QCD confinement scale). 
Top right: the same as top left, but for dark QCD model B (corresponding to low dark QCD confinement scale). 
Bottom left: $C_1^{(\beta)}$ distribution of different kinds of jets with $p_T \in (180\GeV,\,220\GeV)$.
Bottom right: Corresponding ROC curves for discrimination between dark QCD jets and SM QCD gluon-initiated jet.
Here $\beta$ is set to 0.2. 
}
\label{c1}
\end{figure*}

\begin{figure*}[th!]
\centering
\includegraphics[width=0.80\textwidth]{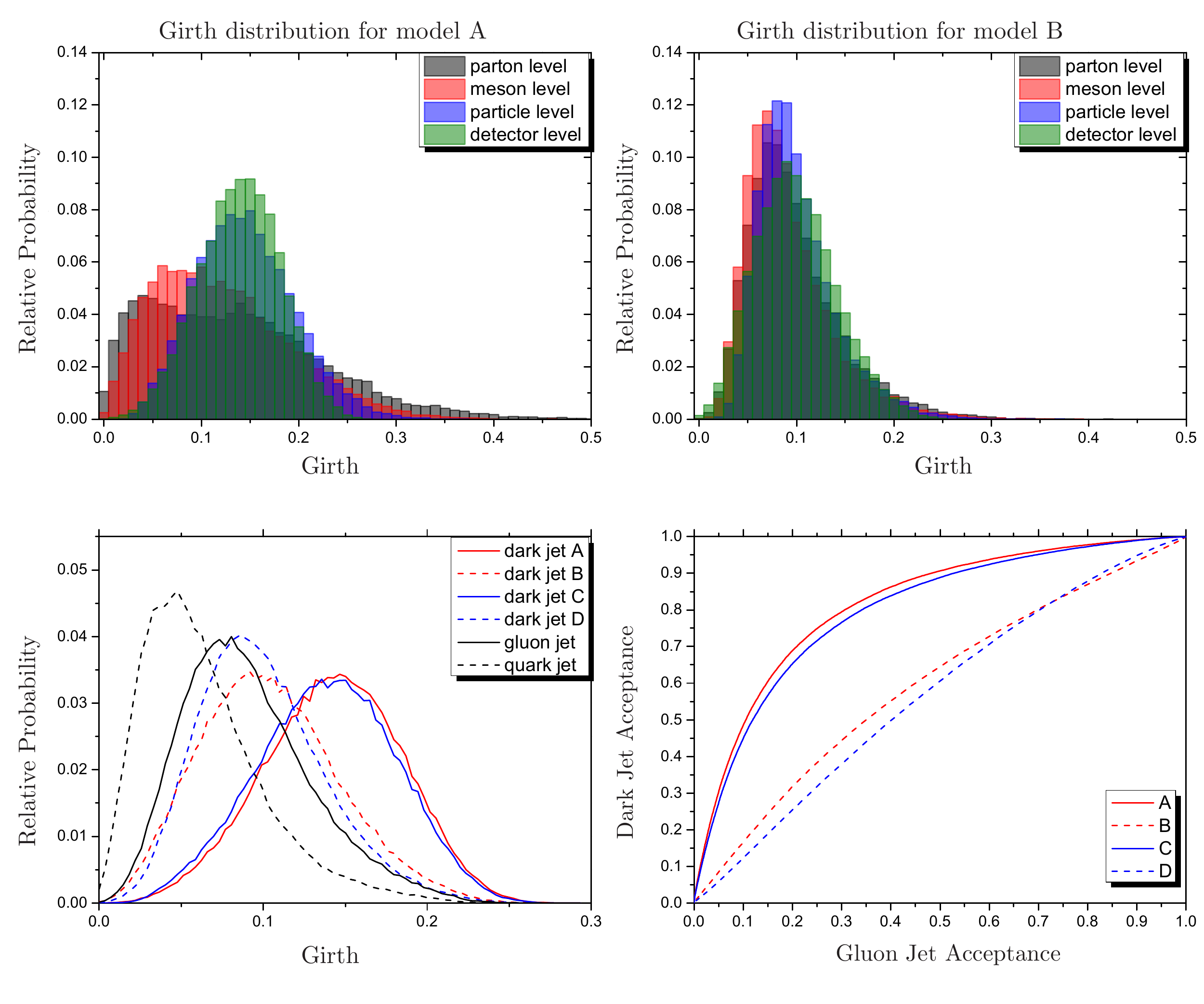}
\caption{Top left: Girth distribution of dark jet with $p_T \in (180\GeV,\,220\GeV)$ at parton level,  meson level, final state particle level, and detector level for dark QCD model A (corresponding to high dark sector confinement scale). 
Top right: Same as top left, but for model B (corresponding to low dark sector confinement scale). 
Bottom left: Girth distribution at detector level of different kinds of jets with $p_T \in (180\GeV,\,220\GeV)$.
Bottom right: Corresponding ROC curves for discrimination between dark QCD jets and SM QCD gluon-initiated jet.
}
\label{girth}
\end{figure*}

There are two more factors that can enhance the discriminant power of $C_1^{(\beta)}$.
Firstly, there is a contribution from non-perturbative fragmentation process. This effect can be estimated by convolving a resummed perturbative distribution with a so-called ``shape" function\,\cite{Gras:2017jty,Korchemsky:1999kt}.
The effect of this convolution is shifting the perturbative distribution of $C_1^{(\beta)}$ to a higher value, and the shift from this non-perturbative process is roughly proportional to the corresponding confinement scale. 
Thus fragmentation process will further separate $C_1^{(\beta)}$ distribution of dark jet and SM QCD jet due to their different confinement scale.

Secondly, when the mass of a dark meson is much larger than SM QCD confinement scale $\Lambda_{QCD}$, the decay of dark mesons inside a jet will strongly affect the distribution of $C_1^{(\beta)}$. This effect can be understood by the following simple estimation.
We consider two nearly collinear dark mesons inside a dark jet, with energy fractions $z_1$,  $z_2$, and distance  $\theta$ between these two dark mesons. $\theta$ should be small because we assume these two dark mesons to be nearly collinear. In this case, contribution from these two mesons to $C_1^{(\beta)}$ is $z_1 z_2\, \theta^{\beta}$.
After both mesons decaying to two SM particles with roughly equal energy, this contribution changes to:
\beq
z_1 z_2\,\theta^{\beta} \to \frac{1}{4}(z_1 + z_2)^2 \left( \frac{m_{\pi_d}}{\overline{p_T}}  \right)^{\beta} .
\eeq
Here $m_{\pi_d}$ is the mass of a dark meson, $\overline{p_T}$ is the average transverse momentum of dark mesons inside a dark jet. 
As we consider a collinear limit between two dark mesons,  an angular distance between dark mesons decay products is approximated as $\left( m_{\pi_d} / \overline{p_T}\right)$.
Thus the mass of a dark meson will increase $C_1^{(\beta)}$ of a dark jet as we consider $\beta >0$.
For a discrimination between quark-initiated jet and a gluon-initiated jet, $\beta$ have been chosen as 0.2\,\cite{Larkoski:2013eya,Bhattacherjee:2016bpy}. 
In this paper, we also follow this choice of $\beta= 0.2$ to distinguish a dark jet from SM QCD jet. 

Simulation results are shown in Fig.\,\ref{c1}. 
Firstly we show $C_1^{(\beta)}$ distributions from parton level to detector level on the top row.
Here, parton level $C_1^{(\beta)}$ means the objects we used to calculate $C_1^{(\beta)}$ is the dark parton after dark shower and before dark hadronization;
meson level $C_1^{(\beta)}$ comes from dark mesons after dark hadronization;
particle level $C_1^{(\beta)}$ comes from all the visible SM particles after dark meson decay;
and detector level $C_1^{(\beta)}$ comes from energy deposits at detector. 
Top left and top right plots are distributions of $C_1^{(\beta)}$ for model A and model B respectively. 
In the top left plot, there is a tall spike at $C_1^{(\beta)}=0$ in the parton level distribution.
This spike comes from a large angle split where one of daughter partons locates outside jet cone. 
At meson level, through convolution with an shape function, this spike at $C_1^{(\beta)}=0$ becomes lower and the distribution is shifted to a higher value. 
Together with this effect, due to the decay of dark mesons, the particle level distribution of $C_1^{(\beta)}$ is pushed to the right side further. 
Finally, the finite resolution of a detector decreases $C_1^{(\beta)}$ to a lower value. 
For model B, due to a weaker coupling and lower confinement scale compared to the case of model A, there is no tall spike of $C_1^{(\beta)}=0$ at parton level distribution, and dark mesons decay pushes up $C_1^{(\beta)}$ only a little. 
In a conclusion, jets from a dark QCD model with a high dark confinement scale jet is easier to tag over SM QCD jets compared to the case of a low dark confinement scale. We also observed that  tagging efficiency is not sensitive to the decay channel of dark meson as $C_1^{(\beta)}$ distribution for A (B) is similar to the distribution of C (D).

\begin{figure*}[ht!]
\includegraphics[width=0.80\textwidth]{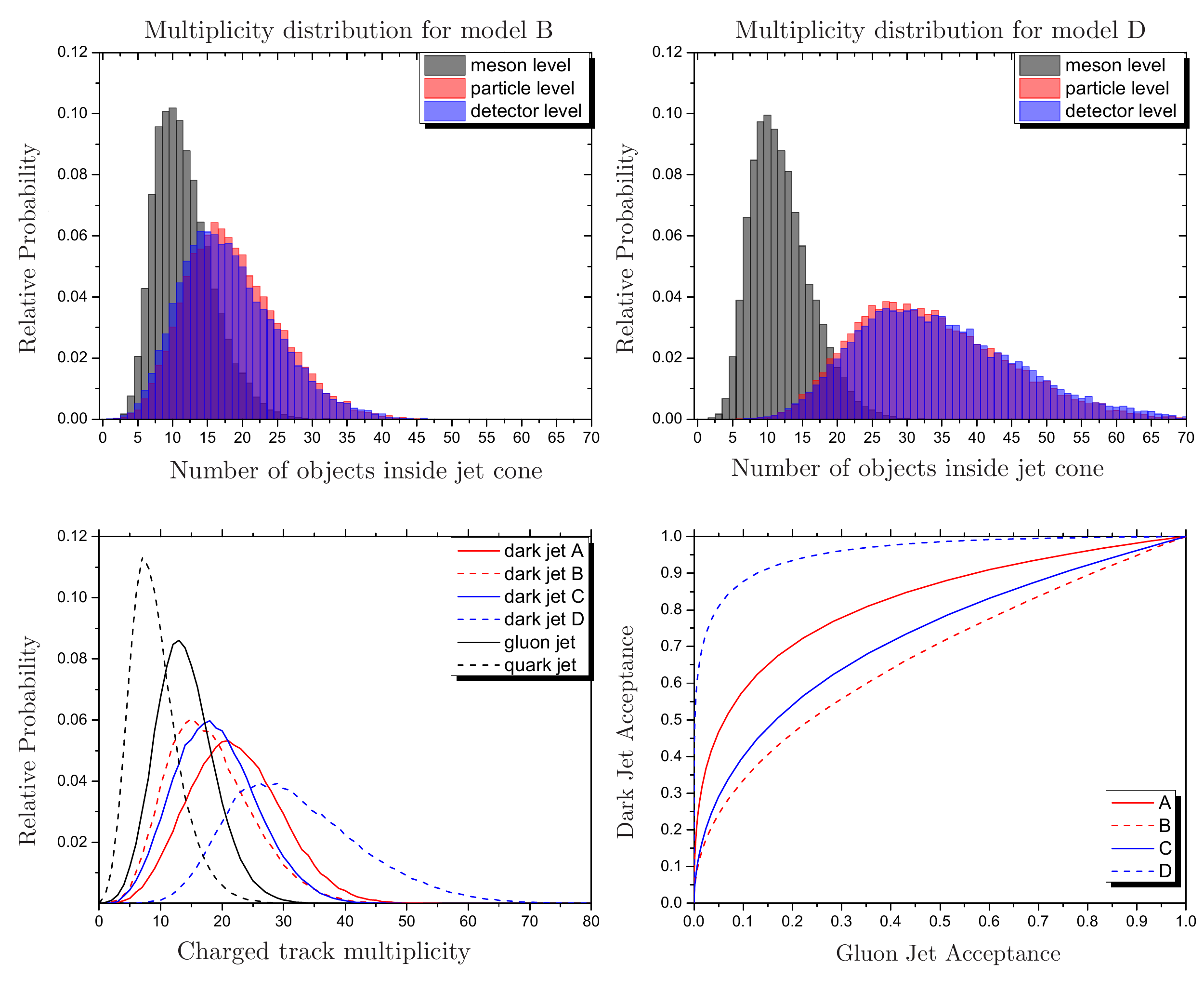}
\caption{Top left: Dark meson multiplicity, charged particle multiplicity, and track multiplicity distribution of dark jet with $p_T \in (180\GeV,\,220\GeV)$ and setting B. 
Top right: Same as top left, but with setting D.
Bottom left: Charged track multiplicity distribution of different kinds of jets with $p_T \in (180\GeV,\,220\GeV)$.
Bottom right: Corresponding ROC curves for discrimination between dark QCD jets and SM QCD gluon-initiated jet.
}
\label{Tnumber}
\end{figure*}

\subsection{Linear Radial Geometric Moment}
Angularity-style variables including jet broadening or width have been studied since LEP period\,\cite{Rakow:1981,Berger:2003iw,Almeida:2008yp,Ellis:2010rwa,Catani:1992jc,Gallicchio:2010dq}.
Here we choose linear radial geometric moment (Girth) to study, which is known as an effective observable in discriminating between quark and gluon jet\,\cite{Gallicchio:2011xq}. Girth is defined as: 
\beq
\textrm{Girth}= \displaystyle{\sum_{i \in J}} \frac{p_{T_i}}{P_{T_J}}\,|r_i|\, ,
\eeq
here $r_i $ is the distance between a component $i$ of the jet and jet axis.
Girth is sensitive to the direction of a jet axis compared to $C_1^{(\beta)}$ which does not require a jet axis.
Here jet axis is defined as the vector sum of all the constituents inside a jet. 

Girth, as a jet width variables, has been analytically analyzed in \cite{Gras:2017jty}.
Here we give a brief description and readers can check more details in \cite{Gras:2017jty} if they are interested. 
At parton level, perturbative calculation shows that quark/gluon jet discrimination ability mainly relies on color factor ration $C_A/C_F$, this is called Casimir scaling. 
For dark jet discrimination, due to a different coupling, the ratio should be replaced by $\alpha_S C_A / \alpha_d C_d $.
Thus one could expect a better discrimination power if $\alpha_d$ is quite different with $\alpha_S$. 
Meson level distribution, as we described in the last subsection, can be obtained by convoluting parton level distribution with a shape function which has a mean value proportional to confinement scale. 
So large $\Lambda_d/\Lambda_{QCD}$ will separate Girth distribution of dark jet and QCD jet further. 
Finally, decay of heavy dark meson will push up Girth value of dark jet. 

Our results from simulations are presented in Fig.\,\ref{girth}. 
In this results, we show the distribution of Girth from model A and model B from parton level to detector level, as we did for $C_1^{(\beta)}$.
Relationship between different levels are as we expected, but the changes are not so much compared to $C_1^{(\beta)}$.
This is because $C_1^{(\beta)}$ is more sensitive to small angular distribution. 
And ,unlike the case of $C_1^{(\beta)}$ variable which needs to have at least two components for non-zero  value, Girth has a non-zero value with even one component\footnote{Here we use the jet axis obtained from detector level to calculate all Girths.}.
Thus a large angle parton splitting doesn't cause a zero point spike in the distribution of Girth as we can find in Fig.\,\ref{c1}.
We conclude that the performance of Girth is dependent on the confinement scale of dark QCD as 
a dark jet from a higher confinement scale is easier to be distinguished than cases from a low confinement scale. With comparison between model A and C (also model B and D) we find that Girth is not sensitive to the different decay channel of a dark meson. And the discriminant ability of Girth is a little weaker than the discriminant ability of $C_1^{(\beta)}$.

\begin{figure*}[ht!]
\includegraphics[width=0.80\textwidth]{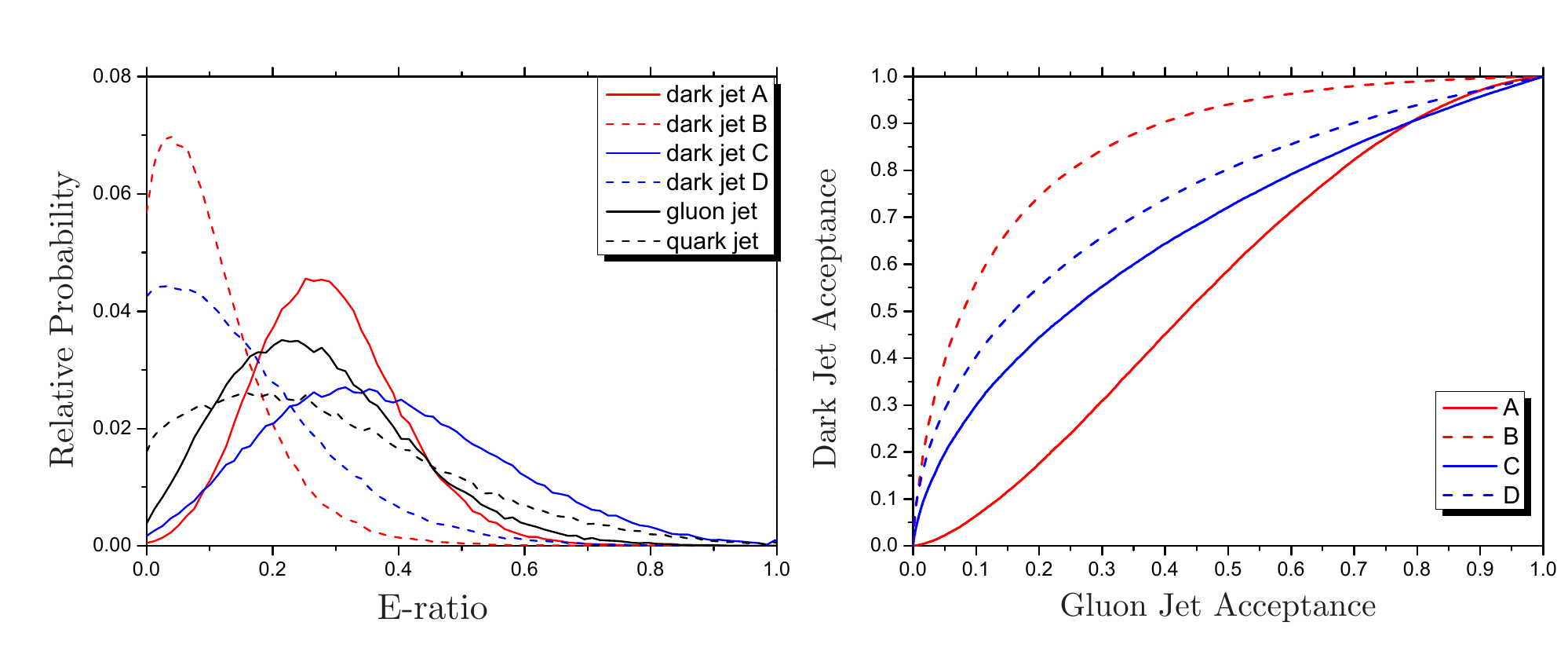}
\caption{
Left: E-ratio distribution of different kinds of jets with $p_T \in (180\GeV,\,220\GeV)$.
Right: Corresponding ROC curves for discrimination between dark QCD jets and SM QCD gluon-initiated jet.
}
\label{Er}
\end{figure*}

\subsection{Charged track multiplicity}
Multiplicity-type variables counting the number of hadrons or tracks inside a jet, turn out to be useful in discriminating different kinds of one-prong jets. 
Among them, charged track multiplicity, due to a high resolution and a trigger efficiency of a track reconstruction at the LHC, is the best discriminant variable among various multiplicity-type variables used in quark and gluon jet discrimination\,\cite{Ackerstaff:1997xg,Gallicchio:2011xq,Gallicchio:2012ez}.
Unlike jet mass or $C_1^{(\beta)}$ which are IRC safe, charged track multiplicity does increase its value through soft and collinear radiations. 
Besides that, it is also closely related to the decay channel of a dark meson. 
So we rely on Monte Carlo simulation results to show its property.

For gluon and quark jets, charged tracks inside them are mainly composed of $\pi^{\pm}$, which come from fragmentation directly. 
But for dark jet, dark mesons coming from dark fragmentation are neutral, and thus tracks inside a dark jet can only come from dark meson decay. 
Due to a very different production mechanism, we expect charged track multiplicity to be a useful discriminant variable.
The amount of tracks produced through dark meson decay is very model dependent. 
For our dark sector setting A, B, C, and D, the average amount of tracks produced by a dark meson are 4.9, 1.6, 4.4, and 2.8.
Dark meson A and C produce more tracks than dark meson B and D. 
This is because the decay of charm mesons and 4.0GeV dark photon generally produce multiple tracks.

Fig.\,\ref{Tnumber} is our simulation results. 
In order to show how the track multiplicity is affected by dark meson's decay channel, we count the amount of dark meson, charged particle, and track with $p_T > 0.5$GeV inside a dark jet, which correspond to meson level, particle level, and detector level respectively in the first row.  
With an identical dark sector setting, dark meson multiplicity distribution for model B and model D are almost the same. 
But different decay channels of dark meson make their track multiplicity quite different. 
Thus compared to dark jet in model B, dark jet in model D is much easier to be discriminated from QCD jet. 
In general, track multiplicity is a better discriminant variable compared with IRC safe variables.

\begin{figure*}[htbp]
\begin{center}
\includegraphics[width=0.8\textwidth]{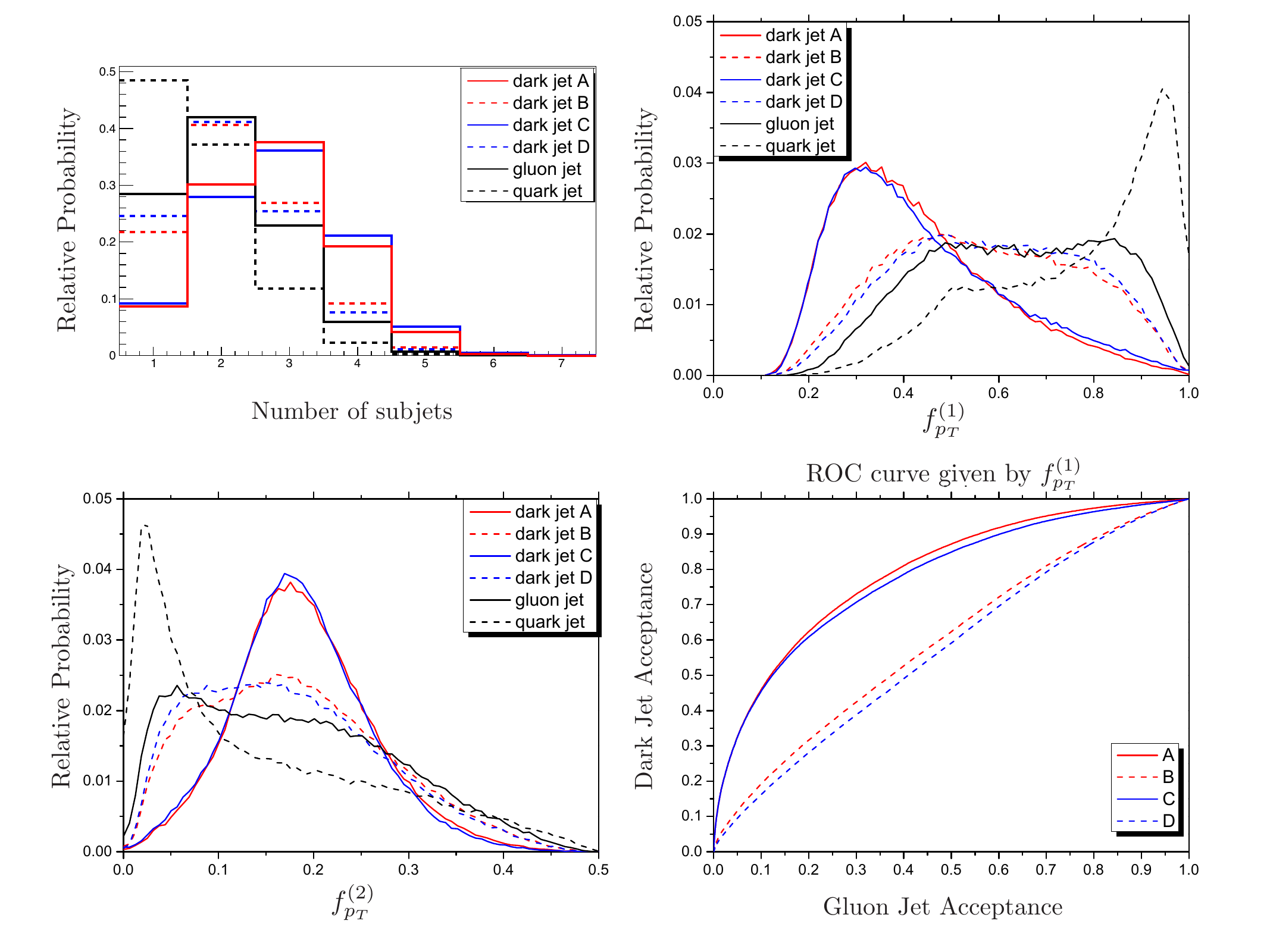}
\caption{Top left: number of sub-jets distribution of different kinds of jets with $p_T \in (180\GeV,\,220\GeV)$.
Top right: hardest sub-jet $p_T$ fraction distribution of different kinds of jets with $p_T \in (180\GeV,\,220\GeV)$.
Bottom left: second hardest sub-jet $p_T$ fraction distribution of different kinds of jet with $p_T \in (180\GeV,\,220\GeV)$.
Bottom right: ROC curves given by the hardest sub-jet $p_T$ fraction $f_{p_T}^{(1)}$.
}
\label{subj}
\end{center}
\end{figure*}

\subsection{Ratio of energy deposits on different calorimeters}
In order to further reflect final states from dark meson's decay, we suggest to utilize a variable which has a dependency on detector's response to different final state particles. 
At the LHC, most of SM particles, except muons and neutrinos, will be stopped by calorimeters and deposit their energy on calorimeters.  
There are two kinds of calorimeters used in the LHC, electromagnetic-calorimeter(ECAL) and hadronic-calorimeter(HCAL). 
Different particles deposit their energy in different calorimeter, as summarized below:
\begin{enumerate}
\item $e^{\pm}$, $\gamma$: deposit energy in ECAL.
\item Hadrons: deposit energy in HCAL. Here hadrons mainly refer to long-lived or stable hadrons like $p$, $n$, $\pi^{\pm}$, and $K^{\pm}$. 
Because hadrons like $\pi^0$ decays to $\gamma\gamma$ before it reach HCAL, and thus deposit energy in ECAL.
\item $\mu^{\pm}$, $\nu$: do not deposit energy in ECAL or HCAL.
\end{enumerate}

Different kinds of jets might have different particle compositions, and thus for a certain jet energy the energy deposited in ECAL might be different. 
So here we define a variable called E-ratio:
\begin{equation}
\text{E-ratio} = \frac{\text{Energy  deposit  on  ECAL }}{\text{Jet's $p_T$}}.
\end{equation}
E-ratio is used to reflect how many energy are carried by $e^{\pm}$, $\gamma$, and $\pi^0$ in a jet, and thus provide more information in addition to track multiplicity.

Particle composition of gluon jet and quark jet are controlled by fragmentation process.
Quarks produced in QCD vacuum are mainly $u$ and $d$. $s$ quark pair are produced with a smaller probability.
Thus background QCD jet are mainly composed of $\pi^0$, $\pi^{\pm}$, and a small quantity of Kaons.
Precise percentage of each kind of particle is given by fragmentation function,  which need to be obtained from data fitting. 
Due to different charge and color factor, E-ratio of gluon jet and quark jet are different\footnote{Actually, because of Parton Distribution Functions (PDF), some quark jets are initiated by $s$ quark.}.
Our simulation result shows that the E-ratio of gluon jet is larger than quark jet. 

For dark jets, dark meson's decay channel controls particle composition. 
Dark meson in model A decays to $c\bar{c}$ pair, and thus most of the energy are carried by $D^{\pm}$ and $D^0$ after fragmentation.
$D^{\pm}$ and $D^0$ decay before reaching calorimeter. 
Decay modes of $D^{\pm}$ and $D^0$ are very complicated, and they mainly decay to $e$, $\mu$, and Kaons through weak interaction. 
Dark jet A turn out to deposit energy on ECAL with a ratio larger than gluon jet. 
Dark meson in model B decays to $s\bar{s}$, and thus the main energy carriers are $K^{\pm}$, $K^{0}_L$, and $K^{0}_S$. 
$K^{\pm}$ and $K^{0}_L$ are long-lived and deposit energy in HCAL. 
Short-lived $K^{0}_S$ decays to 30\% is $\pi^0$ and 70\% $\pi^{\pm}$.
In summary, most of jet energy in model B deposit in HCAL. 

Final decay products of model C is very involved. 
In model C, dark meson decays to dark photon $\gamma'$ pair with a mass 4.0 GeV. 
Main decay modes of 4.0 GeV $\gamma'$ are $d\bar{d}$, $u\bar{u}$, $s\bar{s}$, $c\bar{c}$, $e\bar{e}$, $\mu\bar{\mu}$, and $\tau\bar{\tau}$. 
Thus after all prompt decay, decay products of dark meson is a mixture of $\pi^0$, $\pi^{\pm}$, Kaons, $e$, $\mu$, and $\nu$. 
In all of our models, dark jet C deposit most energy in ECAL.
Finally, dark meson D decays to $\gamma'$ pair with a mass 0.7 GeV. 
0.7GeV $\gamma'$ decays to $e\bar{e}$ by a probability of 15\%, and other decay products are $\mu$ and $\pi^{\pm}$.
So dark jet D does not deposit much energy in ECAL and its E-ratio is small.

Distribution of E-ratio is shown in Fig.\,\ref{Er}. As we expected, E-ratio distribution of model B and D are small compared with other jets. and corresponding ROC curve shows a good discriminant performance. 
While for model A and model C, this variable is not so useful.

\begin{figure*}[htbp]
\begin{center}
\includegraphics[width=0.7\textwidth]{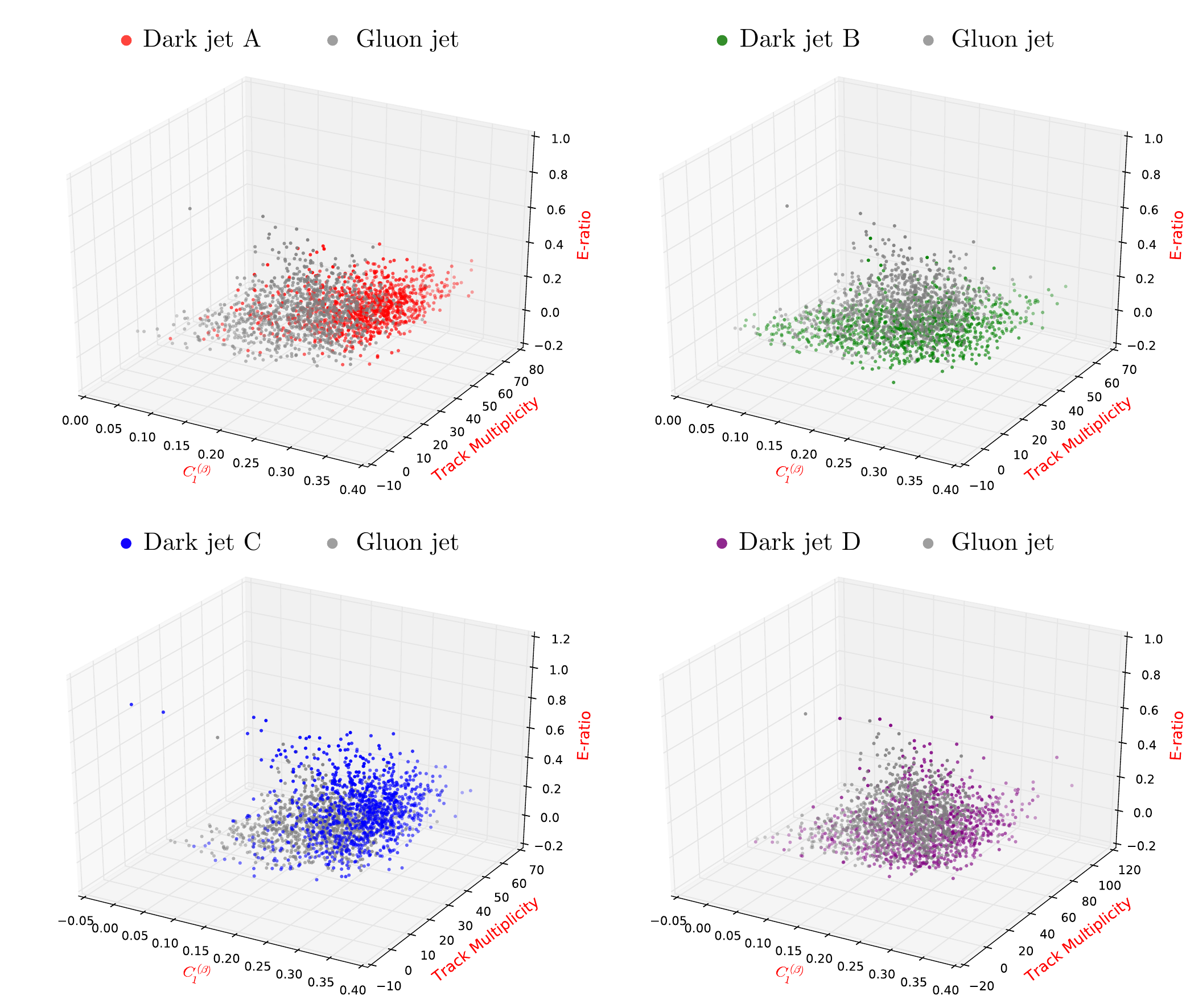}
\caption{Variables distribution in $\{C_1^{(\beta)},\,\textrm{E-ratio, Track Multiplicity}\}$ space. 
Red, green, blue, purple, and grey points are samples of dark jet A, dark jet B, dark jet C, dark jet A, and gluon jet respectively.
}
\label{3D}
\end{center}
\end{figure*}

\subsection{Sub-jet}
Properties of an one-prong jet can also be revealed by measuring observables associated with smaller sub-jets inside it. 
Because different kinds of jets have different energy profiles on a transverse plane. 
For example, most of the energy of quark jet concentrate on a small central region, while the energy of gluon jet will spread to a larger area\cite{Gallicchio:2011xq}. 
Here we define a sub-jet by re-clustering constituents of an original jet with anti-kt algorithm and a jet radius $R = 0.1$.   
We require the $p_T$ of these sub-jets to be larger than 5\% of the original jet's $p_T$.
Here we define $f_{p_T}^{(i)}$ as $p_T$ of $i$-th hardest sub-jet divided by $p_T$ of an original jet:
\begin{eqnarray}
f_{p_T}^{(i)} = \frac{\text{$p_T$ of $i$-th hardest sub-jet}}{\text{original jet's $p_T$}}.
\end{eqnarray}
Three variables are used here: 1)\,the number of sub-jets, 2)\,$p_T$ fraction carried by the hardest sub-jet $f_{p_T}^{(1)}$, and 3)\,$p_T$ fraction carried by the second hardest sub-jet $f_{p_T}^{(2)}$.

Simulation results are shown in Fig.~\ref{subj}. 
Those distributions show clear physical meaning. 
QCD quark jet, with a small coupling and color factor, can only trigger large angle shower with a quite low probability. 
Hence there is a huge possibility for quark jet to concentrate most of its energy in a tiny cone with a radius smaller than 0.1.
Due to a larger color factor, QCD gluon jet is "broad" compared to "narrow" quark jet, which means the energy of gluon jet distribute on a larger area and it's more likely to have more sub-jets inside gluon jet.
For dark jet, through a larger coupling, they become even more broader and there are more sub-jets inside it. 
$p_T$ fraction of sub-jets are natural expectation of such argument. 
Among these 3 variables, $p_T$ fraction of the hardest sub-jet $f_{p_T}^{(1)}$ shows the best discriminant ability. 
Similar to $C_1^{(\beta)}$, Girth, and jet mass, this variable is useful when dark confinement scale is high.

\begin{figure*}[ht!]
\includegraphics[width=0.80\textwidth]{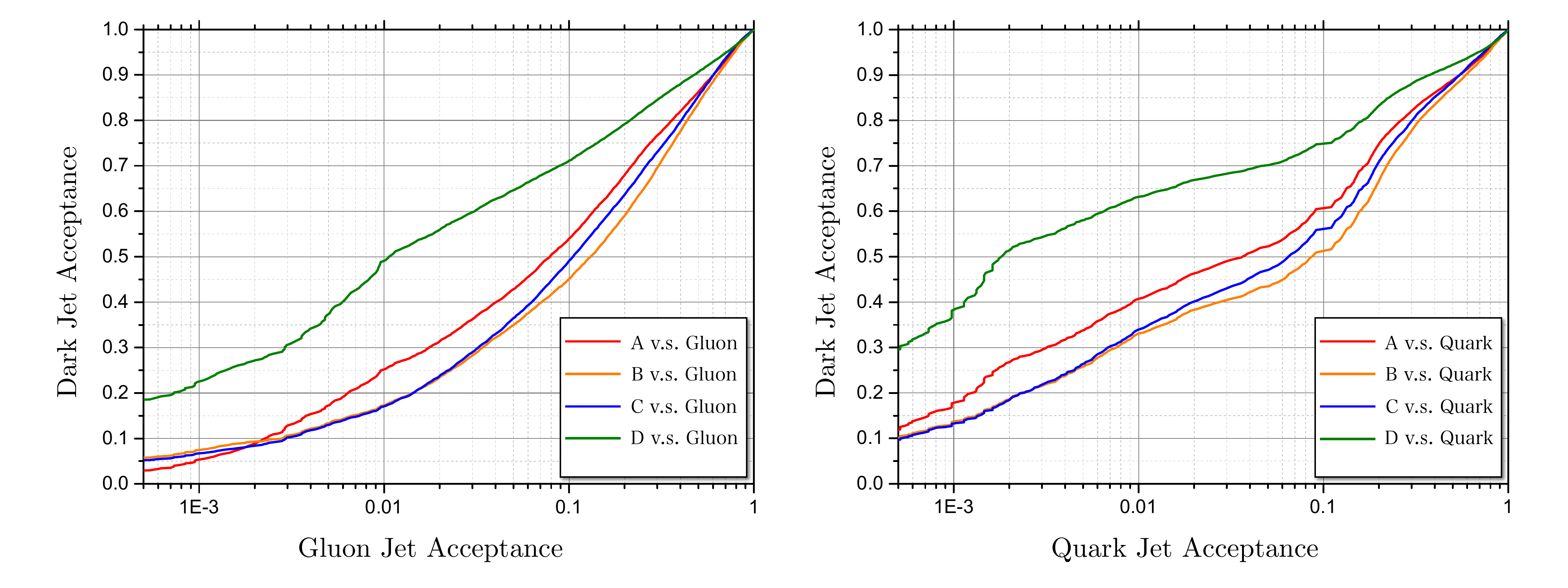}
\caption{
ROC curves obtained by cutting on BDT score. 
Left: dark jet vs. gluon jet ROC curves for all four models, with $p_T \in (100\GeV,\,450\GeV)$.
Right: dark jet vs. quark jet ROC curves for all four models, with $p_T \in (100\GeV,\,450\GeV)$.
}
\label{BDT}
\end{figure*}

\subsection{Combine multiple variables}
\label{combine_BDT}

\begin{table*}[t!]
\begin{center}
\begin{tabular}{|c|c|c|c|c|}\hline   & Model A & Model B & Model C & Model D \\\hline
 Best discriminant variable & $C_1^{(\beta)}$ & E-ratio & $C_1^{(\beta)}$ & Track Multiplicity \\\hline 
 \end{tabular}
 \caption{Best discriminant variable for different models.}
  \label{bestv}
\end{center}
\end{table*}

One can estimate the discriminant performance of a variable by simply counting the area under the ROC curve, and a bigger area means a better discriminant ability. 
We show the best discriminant variable for each model in Tab.\ref{bestv}. 
Discriminant performance can be maximized by combining multiple jet-substructure variables.
To start our multiple variables analysis, we consider a set of variables consisting of the best discriminant variables: $\{C_1^{(\beta)},\,\textrm{E-ratio, Track Multiplicity}\}$. 

In principle, a classification can be made by a hyperplane in $\{C_1^{(\beta)},\,\textrm{E-ratio, Track Multiplicity}\}$ space, but finding such a hyperplane is often difficult. 
For illustration, in Fig.~\ref{3D}, we show sample distribution of different dark jets and gluon jet in $\{C_1^{(\beta)},\,\textrm{E-ratio, Track Multiplicity}\}$ space.
It can be seen that the distribution of dark jets are intertwined with the distribution of gluon jet, and different dark jets populate different regions. 
Due to these complexities, in this work we use Boosted Decision Tree (BDT)~\cite{Roe:2004na} in TMVA-Toolkit~\cite{Hocker:2007ht} to do multiple variables analysis. 

BDT need to be trained by sufficient amount of samples that represent the signal and background characteristics. 
For both signal and background samples, we require their jets $p_\text{T}$ to evenly distribute from $100\GeV$ to $450\GeV$, and pseudo rapidity $\eta\,\in(-2.5,2.5)$.
Our signal sample contains 100k events, with 17\%, 34\%, 45\%, and 4\% of them coming from setting A, B, C, and D respectively\footnote{This sample ratio choice comes from feasibility requirement. We found dark jets A and D are easier to distinguish than dark jets B and C. Thus in order to obtain a BDT that is useful to all kinds of dark jets, we need to increase the proportion of events from model B and C in signal sample. After several attempts, we found that the sample ratio used here gives us a BDT with good feasibility.}. 
Our background sample contains 30k events, and all of them are gluon jets. 
Because the distributions in previous subsections indicate that discrimination between gluon jet and dark jet should be harder than discrimination between quark jet and dark jet.
We use 500 decision trees, choose minimum in leaf node as 2.5\%, and set maximum depth as 3. 
To avoid overtraining, half of the events are chosen as test events and Kolmogorov-Smirnov test is required to be larger than 0.01. 

After training, BDT can map an event with a certain $\{C_1^{(\beta)},\,\textrm{E-ratio, Track Multiplicity}\}$ value, to a BDT score. 
A signal-like event tends to get high BDT score, and a background-like event tends to get low BDT score. 
Thus a discrimination can be performed by simply cutting on BDT score. 
In Fig.\,\ref{BDT} we show the ROC curve of dark jet and gluon jet discrimination obtained by cutting on BDT score. 
As one can expect, dark jet D is the easiest to distinguish, because dark jet D generally produce much more tracks than gluon jet. 
And the difficulty of distinguishing dark jet A, B, and C from background are close to each other. 
For comparison we also present ROC curve for dark jet vs. quark jet in Fig.\,\ref{BDT}. 
It shows that the BDT trained by dark jet and gluon jet can also be used to do discrimination between dark jet and quark jet.
And as we expect, distinguishing dark jet from quark jet is easier than distinguishing dark jet from gluon jet.
The trained BDT we used here is also workable to other kinds of dark jets. 
More discussion on feasibility can be found in Appendix B.

Generally, if we use more variables in BDT we might get a better discriminant performance.
But if a variable is strongly correlated with other variables, then this variable will be redundant and can not provide more information about a jet. 
In our training we also consider an extended set of variables which contains 8 variables: \{$m_J/p_T$, $C_1^{(\beta)}$, Girth, Track Multiplicity, E-ratio, number of sub-jets, $f_{p_T}^{(1)}$, $f_{p_T}^{(2)}$\}. 
But the improvement we can get by including 8 variables are negligible. 
It indicates that the main feature of a dark jet is described by $C_1^{(\beta)},\,\textrm{E-ratio, and Track Multiplicity}$.

The BDT we trained in this section is feasible to different kinds of dark jets, but it might have lost some of the characteristics of the samples. 
Considering more than one BDT may help to comprehensively show the nature of the sample.
For a discussion of 2 BDTs, see Appendix C.

\section{An example at LHC}
\label{sec:example}
\begin{figure*}[t!]
\begin{center}
\begin{tabular}{cc}
\includegraphics[width=0.80\textwidth]{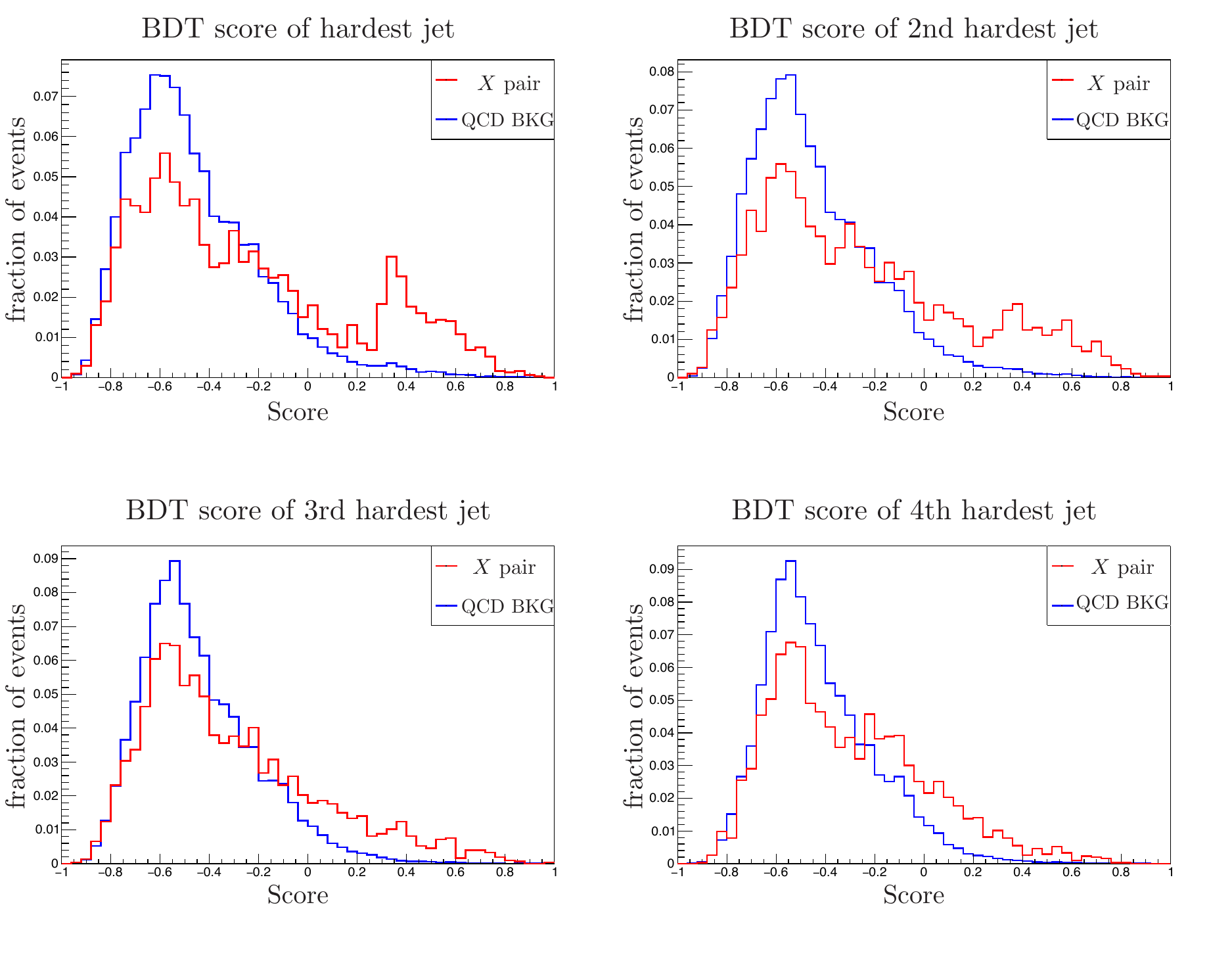} 
\end{tabular}
\end{center}
\caption{\label{BDT_score} BDT score distributions of 4 leading jets of signal and background. Events used here are required to have at least 4 jets with $p_\text{T}> 120$GeV and $|\eta|<2.4$. BDT score is normalized to region [-1,+1]. For signal process, here we consider $M_X = 400$GeV for illustration. 
}
\end{figure*}
In this section, we present bi-fundamental scalar $X$ search at LHC as an example to utilize our dark jet tagging method.  
As $X$ is charged under both the SM $SU(3)$ and dark $SU(3)$, pair of $X$ particle can be produced at LHC through QCD process. 
Once a mediator $X$ is produced, it decays into a SM quark and a dark quark, which evolves to a QCD jet and a dark jet respectively. 
If the decay length of a dark meson is around $\mathcal{O}(10) \sim \mathcal{O}(100)$ $mm$,  a dark jet will leave displaced vertices in detector. 
By counting the number of displaced vertices, one can obtain robust limit on the mass of mediator particle $X$\cite{Schwaller:2015gea,CMS:2018mls}.
But if the decay length of a dark meson is shorter than $1mm$, analyses with displaced vertices will lose sensitivity. 
In this section we will show that tagging dark jet with jet-substructures can be used to enhance a search sensitivity when dark mesons decay promptly.

\begin{table*}[t!]
\begin{center}\begin{tabular}{|c|c|c|c|c|c|}
\hline  & BKG & $M_X = 300$GeV & $M_X = 500$GeV & $M_X = 700$GeV  & $M_X = 900$GeV \\
\hline  \tabincell{c}{ 4 jets with $|\eta|<2.4$ \\ and $p_\text{T}> 120\GeV$ } & $2.46\times 10^7$ & 70,701  & 21,723  & 5,127  & 1,267  \\
\hline  $\Delta R_\text{min}$ selection  & $3.96\times 10^6$ & 20,998  & 5,946  & 1,631  & 499.6  \\
\hline  \tabincell{c}{ $\mathcal{A} < 0.05$ and \\ $|\cos\theta^{\ast}| < 0.3$ } & 154,750 & 2,246  & 529.7  & 120.7  & 30.6  \\
\hline  Dark jets $\geqslant$ 1 &  2,240  & 403.0 & 130.2 & 41.6  & 12.3  \\
\hline Dark jets $\geqslant$ 2 & 19.9 & 30.1 & 15.3 & 5.0  & 1.8  \\
\hline Significance & - & 4.03 & 2.05 & 0.68  & 0.25  \\
\hline \end{tabular} \caption{The number of signal events and background events after applying cut. Luminosity is 36.7 fb$^{-1}$ and central energy is 13 TeV. Background events number after inclusive selection have been normalized to the data observed in ATLAS report. }
\label{Cutflow}
\end{center}
\end{table*}

We consider dark sector setting A in Tab.\ref{Models} as an benchmark for the LHC study.
Our analysis is based on the search for pair-produced resonances in four-jet final states on ATLAS\,\cite{Aaboud:2017nmi}.
Here we briefly describe the cut flow used in ATLAS report\,\cite{Aaboud:2017nmi}:
\begin{itemize}
\item Events are required to have at least 4 jets with $p_\text{T}> 120\GeV$ and $|\eta|<2.4$. 
\item These 4 jets are paired by minimizing $\Delta R_\text{min} = \sum_{i=1,2} |\Delta R_i -1 | $, with $\Delta R_i$ the angular distance between two jets in a pair.  
\item Define $\bar m$ as the average of the invariant masses of these two jets pair as $\bar m=\frac{1}{2}(m_1+m_2)$. 
Here $m_1$ and $m_2$ are the invariant masses of two resonances. 
We veto events of large angular separation  
\beq
\nonumber \Delta R_\text{min} > -0.002  (\bar m- 225) + 0.72\,
\eeq
for the case of $\bar m<225\GeV$. If $\bar m\ge225\GeV$, we discard events with
\beq
 \Delta R_\text{min} > 0.0013 (\bar m- 225) + 0.72.
\eeq
\item Boosting the system of these two resonances (two jets pairs) to their centre-of-mass frame. $\cos\theta^{\ast}$ 
is defined as the cosine of the angle between one of the resonance and the beam-line in the centre-of-mass frame.
The mass asymmetry $\mathcal{A}$ is defined as:
\begin{eqnarray}
\mathcal{A} =  \frac{|m_1-m_2|}{m_1+m_2},
\end{eqnarray}

Events are selected by requiring $\mathcal{A} < 0.05$ and $|\cos\theta^{\ast}| < 0.3$. This cut flow defines the inclusive signal region (SR) selection.
\end{itemize}

This analysis utilizes kinetic information of final state jets, which are $p_\text{T}$, $\eta$, and $\phi$. 
However, as we have presented in section III, one can get more information by looking inside a jet.
If the resonance is the mediator particle $X$, there will be dark jets in the final state. 
So, by tagging dark jets, search sensitivity can be enhanced. 
Here we use the trained BDT we obtained in last section to compute BDT scores of 4 leading jets in the final state. 
Thus dark jet can be tagged by cutting on BDT score. Similar method have been performed in SUSY study\,\cite{Bhattacherjee:2016bpy}.

\begin{figure}[th!]
\begin{center}
\begin{tabular}{cc}
\includegraphics[width=0.5\textwidth]{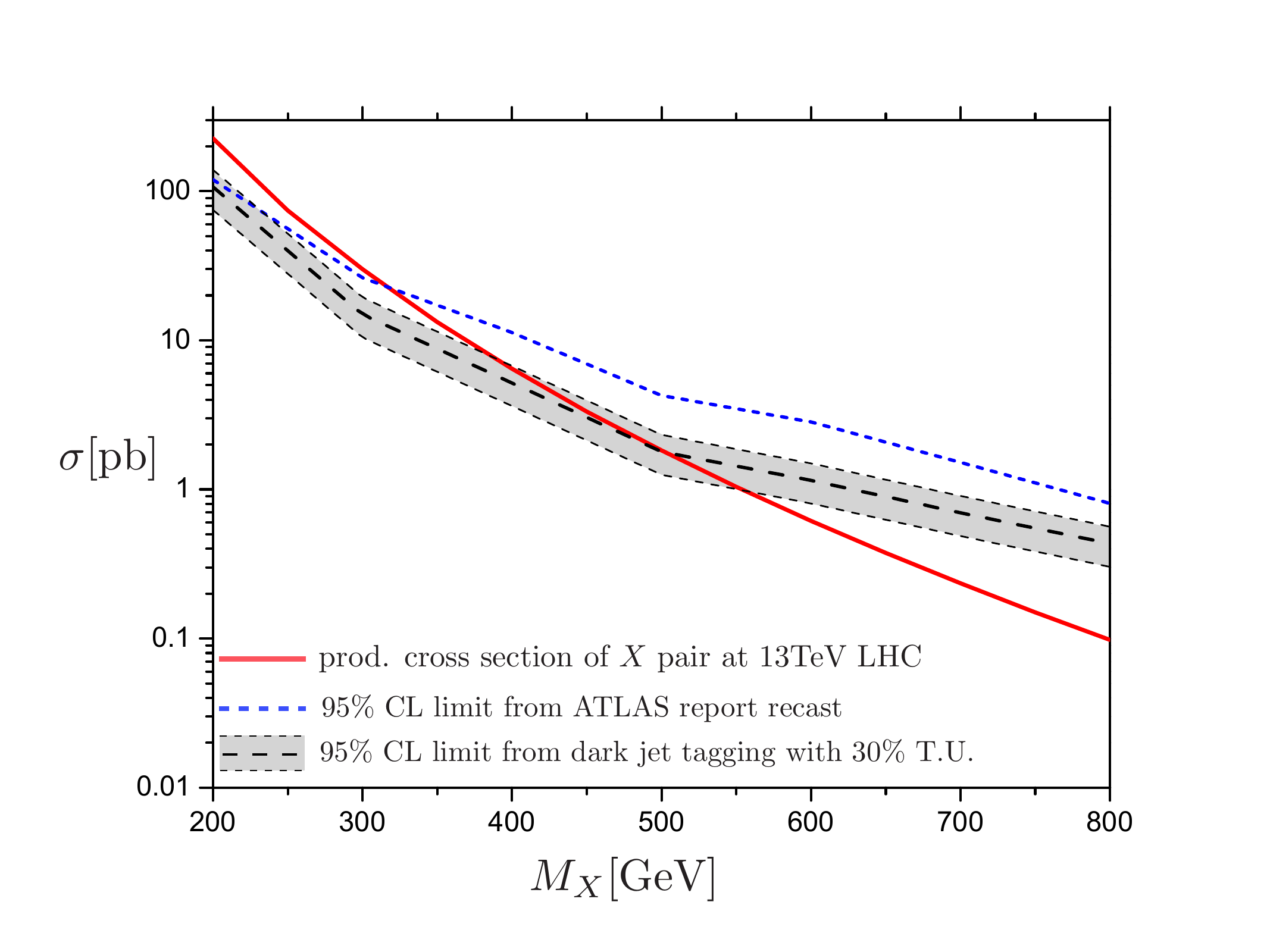} 
\end{tabular}
\end{center}
\caption{\label{exclu} The 95$\%$ CL upper limit on the production cross section of $X$ pair, with $X$ decays to a SM quark and a dark quark. Red line is the production cross section of $X$ pair at 13TeV LHC. Blue dashed line is the up limit obtained by using the cut flow in ATLAS report\cite{Aaboud:2017nmi}. Black dashed line is the up limit obtained by using our dark jet tagging method. Gray band reflect theoretical uncertainty (T.U.) from dark jet tagging. 
}
\end{figure}

Backgrounds from SM QCD processes and signal events from $X$ pair production are generated by \pythia. 
For background simulation, we generate more than 1 billion events, and the events number after inclusive cut is normalized to the data observed in ATLAS report \cite{Aaboud:2017nmi}.
The production cross section of $X$ pair is the production cross section of stop pair multiplied by 3~\cite{stop_Xs}, because X is also charged under a dark $SU(3)$ gauge group. 
In Fig.\ref{BDT_score} we show the BDT score distributions of 4 leading jets for both background and signal.
It can be seen that the BDT score of our signal is larger than the BDT score of QCD background. 
Thus we can suppress QCD background by cutting on the BDT scores. 
Here we define a jet with BDT score larger than 0.483\footnote{This number is chosen for maximizing exclusion limit.} to be tagged as dark jet, and we require at least 2 dark jets in final state.
In Tab.\ref{Cutflow} we list the cutflow for both background and signal. 
After inclusive selection, background events number is 3 orders of magnitude larger than signal events number. 
By requiring at least two dark jets in the final state, we suppress background by 4 orders of magnitude, and thus background and signal become comparable.
Significances are estimated by $\frac{S}{\sqrt{B+\epsilon^2 B^2 }}$. Here $S$ and $B$ are events number of signal and background respectively, and we assume systematic uncertainty $\epsilon$ to be 30$\%$ for a conservative estimation. 

We give a $95\%$ confidence level exclusion limit on the cross-section of X pair production in Fig.\,\ref{exclu}.  
Theoretical uncertainty (T.U.) from dark jet tagging is estimated by varying renormalization-scale $\mu$ in dark parton shower process. 
See Appendix A for a detailed discussion about theoretical uncertainty. 
In order to compare with the search method without dark jet tagging, in Fig.\ref{exclu} we also show the up limit obtained by doing a recast of ATLAS report \cite{Aaboud:2017nmi}.
In report \cite{Aaboud:2017nmi}, after inclusive cut, several mass window are designed to further increase search sensitivity. 
For a certain resonance mass, average mass $m_\textrm{avg}$ is required to be located in a narrow region around it. 
While due to a strong shower in dark sector, the average mass obtained by 4 final state leading jets distribute in a broad mass region. 
Thus the mass window cut discard too many signal events and result in a low sensitivity. 
Fig.\ref{exclu} shows that the exclusion limit from ATLAS report recast is weaker than the limit from our dark jet tagging method.

\section{Conclusion}
Dark sector charged under a confined $SU(N_d)$ provides composite states and attractive phenomenologies.
At colliders, such model can produce jet-like signal (called ``dark jet''), some of which may not be tagged by distinct or exotic signatures including missing energy or displaced vertex. 
In this work, inspired by the success of quark/gluon jet discrimination, we try to distinguish dark jet from background SM QCD-jet by using jet sub-structure variables. 
A series of jet-substructure variables, like the jet mass, $C_1^{(\beta)}$, or track multiplicity, are studied in this work. 
We use simulated events to exhibit discriminant ability of all these jet-substructure variables.
A BDT is trained as classifier to discriminate dark jet and QCD jet. 
For all of our dark sector settings, we can use this trained BDT to exclude 99\% background gluon jets with more than 15\% signal dark jet reserved, or exclude 99\% background quark jet with more than 30\% signal dark jet reserved. 
A concrete example is used to show that our dark jet tagging method can enhance the search sensitivity of models which produce dark jets in final state. 
Theoretical uncertainty from Monte Carlo simulation is also discussed, which shows that our method is robust against theoretical uncertainty.

\section*{Acknowledgements}
Authors appreciate Dr.\,Sung Hak Lim for discussions at the earlier stage of this project.
MZ appreciate helps from Pythia authors, Torbjorn Sjostrand, Stephen Mrenna, and Peter Skands.
MZ thanks Yasuhito Sakaki, Jinmian Li, and Zhuoni Qian for useful discussion with them. 
This work is supported by IBS under the project code, IBS-R018-D1. MP is supported the Basic Science Research Program through the National Research Foundation of Korea Research Grant NRF-2017R1C1B5075677 and NRF-2018R1C1B6006572.

\appendix
\section{Uncertainty Discussion}
The discriminant ability shown in section~\ref{sec:example} might be quite sensitive to theoretical uncertainty (T.U.) from Monte Carlo event generator. 
In analyses of quark-gluon jet tagging, one can tune parameters in the Monte Carlo event generator from real data to reduce systematics and enhance predictability.
Generally, one can simulate quark jet very well with public event generators. 
And for gluon jet, it is also known that the real distribution lies in between Pythia and Herwig~\cite{Bahr:2008pv} expectation.
More information can be found in recent review~\cite{Gras:2017jty}.

\begin{figure}[htbp]
\begin{center}
\includegraphics[width=0.45\textwidth]{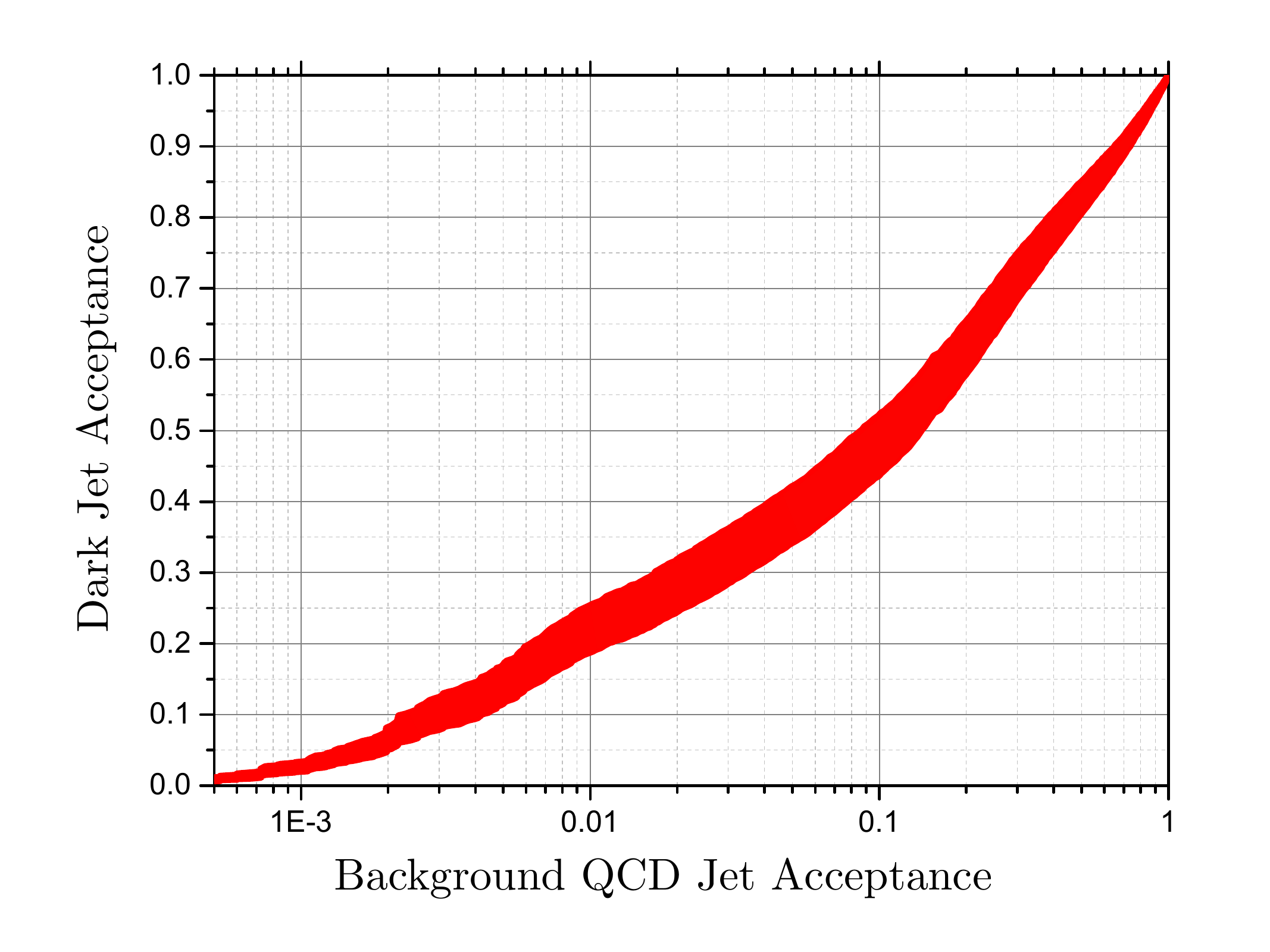}
\caption{
ROC discriminant curve of dark jet vs. SM QCD jet, with renormalization scale $\mu$ varying from $0.5p_\bot$ to $2.0p_\bot$. 
Here we use dark sector setting A in Tab.\ref{Models}. 
}
\end{center}
\label{unc}
\end{figure}

But for a dark jet, we can not estimate systematics as we don't have real data of dark jet. Thus parameters in simulating dark QCD hadronization and showering leaves unfixed systematics in our analyses.
On top of this difficulty, as we don't have various Monte Carlo generators for dark jet simulation except \pythia, we don't have a choice to  compare different event generator to get an estimation about uncertainty depending on different showering and hadronization schemes.
Alternatively, we do some simple estimation in this work. 
Changing renormalization scale in parton shower process has been proved to be a good method to estimate theoretical uncertainty in Pythia~\cite{Mrenna:2016sih}.
So following this method, we also rescale the renormalization scale $\mu$ in dark sector shower process.

For background and signal events simulation, we use the same setting as we used in section~\ref{sec:example}.
But we simulate signal events several times with different renormalization scale $\mu$, varying from $0.5p_\bot$ to $2.0p_\bot$.
Here $p_\bot$ is the order parameter used in \pythia shower process, and $\mu = p_\bot$ is a default set in \pythia.
So after utilizing the trained BDT to do discrimination we obtain several ROC curves. 
Results are shown in Fig.~\ref{unc}. 
In the region where the acceptance of background QCD jet is around 1$\%$, the acceptance of signal dark jet changes from 19$\%$ to 26$\%$. 
So if we want to tag one dark jet in final state, corresponding theoretical uncertainty is about 15$\%$. 
In section~\ref{sec:example} we want to tag two dark jets in final state, thus the theoretical uncertainty is about 30$\%$. 


\section{Feasibility Discussion}

\begin{table*}[t!]
\begin{center}\begin{tabular}{|c|c|c|c|c|c|c|p{2.7cm}|p{3cm}|c|}
\hline       &
 \multirow{ 2}{*}{$N_d$} & \multirow{ 2}{*}{$n_f$} & $\Lambda_d$ &  $\tilde{m}_{q'}$ & $m_{\pi_d}$ & $m_{\rho_d}$ & 
 \multirow{ 2}{*}{$\pi_d$ Decay Mode} & \multirow{ 2}{*}{$\rho_d$ Decay Mode} \\
  &  & & (GeV) &  (GeV) & (GeV) & (GeV) &  & \\
\hline$E$ &3&3&8&  10  &5&25&$\pi_d \to c \bar{c}$&$\rho_d \to \pi_d \pi_d$ \\
\hline \multirow{ 2}{*}{$F$} &\multirow{ 2}{*}{3}&\multirow{ 2}{*}{3}&\multirow{ 2}{*}{5}& \multirow{ 2}{*}{6} &\multirow{ 2}{*}{5}&\multirow{ 2}{*}{14.33}&$\pi_d \to \gamma' \gamma' $ {with $m_{\gamma'} = 0.8\GeV$} &\multirow{ 2}{*}{$\rho_d \to \pi_d \pi_d$}  \\
\hline\end{tabular}
\caption{Benchmark models used to show the feasibility of the BDT we trained in section III.}
\label{Models_EF}
\end{center}
\end{table*}

\begin{figure*}[ht!]
\includegraphics[width=1.0\textwidth]{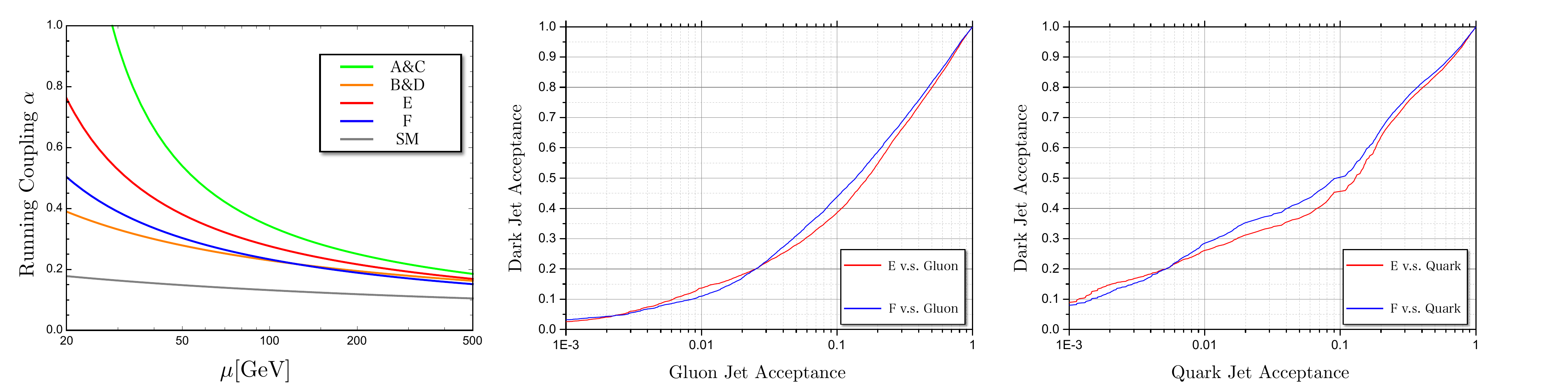}
\caption{
Left: QCD coupling running in dark sector and SM QCD.
Middle: dark jet vs. gluon jet ROC curves for benchmark points E and F, with $p_T \in (100\GeV,\,450\GeV)$.
Right: dark jet vs. quark jet ROC curves for benchmark points E and F, with $p_T \in (100\GeV,\,450\GeV)$.
}
\label{BDT_EF}
\end{figure*}

In order to further test the feasibility of the BDT trained in section III, we consider another two benchmark models E and F. 
Detailed setting can be found in Tab.\ref{Models_EF}.
We choose these two settings because they have confinement scales different with setting A and B. 
In the left plot of Fig.\ref{BDT_EF} we present the coupling running of E and F. 
It can be seen that the running coupling of E and F locate between setting A and B. 
For model E and F we generate 50k events with jet $p_T$ ranging from 100 GeV to 450 GeV. 
As we did in section III, we use the trained BDT to map the value of $\{C_1^{(\beta)},\,\textrm{E-ratio, Track Multiplicity}\}$ to a BDT score. 
By cutting on BDT score, we obtain ROC curves, which are shown in Fig.\ref{BDT_EF}. 
The discriminant performance is quite good.
For model E and F, we can exclude 99\% background gluon jets with more than 11\% signal dark jet reserved, or exclude 99\% background quark jet with more than 26\% signal dark jet reserved. 

\begin{figure*}[ht!]
\includegraphics[width=1.0\textwidth]{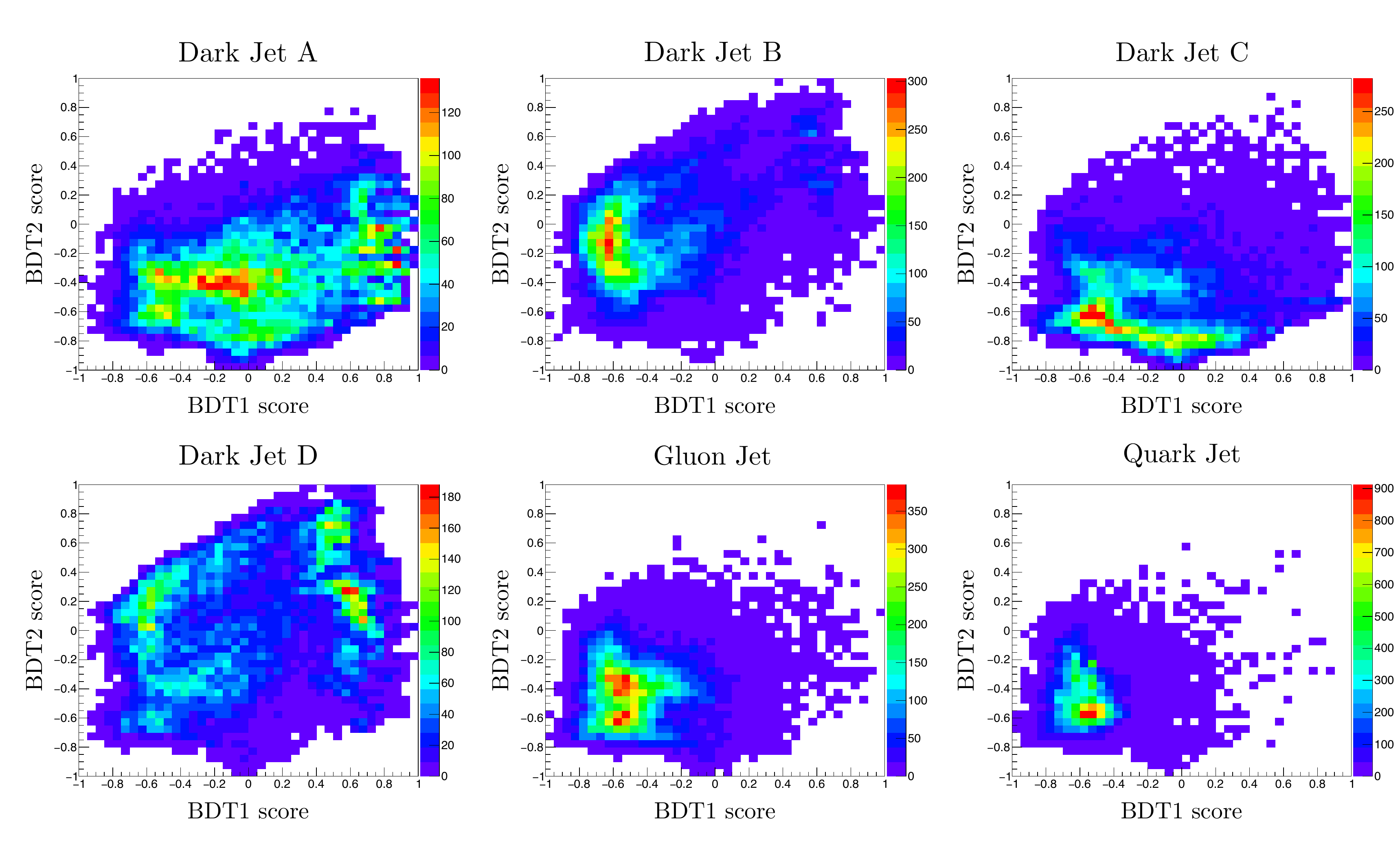}
\caption{Distributions on ``BDT1 score''-``BDT2 score'' plane of different kinds of dark jets and background QCD jets. Distribution in each plot are obtained from 30k events. 
}
\label{BDT_plane}
\end{figure*}

Here we emphasize that we train BDT with samples from benchmark points A, B, C, and D to cover generic features of dark jet.
More specifically, our BDT captures a feature of a large coupling by studying sample A and it also learns different decay modes from C.
To prepare small coupling scenarios, we provide a chance of learning with benchmark point B and D for each case.
As we test its performance on different parameter points (E and F), our method can be applied to wide range of parameters in dark QCD physics.

\section{Discrimination by 2 BDTs}
In Sec.~\ref{combine_BDT} we use a mixture of the events from setting A, B, C, and D as signal sample to train a BDT,
and this BDT shows a good feasibility in distinguishing different kinds of dark jets from QCD background. 
But such a BDT might lose some unique characteristics of a certain kind of dark jets. 
This problem can be alleviated if we use two BDTs. 
In this appendix we consider a BDT1, which is trained by choosing dark jet A as signal sample, and a BDT2, which is trained by choosing dark jet B as signal sample.  
Thus for each jet we will get two BDT scores, and more information about a jet can be reflected. 
We can not easily obtain a ROC curve by cutting on BDT scores in this case. 
So instead of presenting a ROC curve, we simply show the distribution of different kinds of jets on the BDT scores plane. 
Fig.~\ref{BDT_plane} is our result. 
It shows that both gluon and quark jets concentrate on lower left quarter, but dark jets can spread to larger region. 
Dark jet A and Dark jet D populate more than half of the BDT score plane.
Dark jet B and Dark jet C also show distributions which are different with QCD background. 
Discrimination in this case can be performed by some image identification technique.
But a study on utilizing image identification is beyond the scope of current work, and we stop our discussion here.


\end{document}